**Tool and tutor?**
**Experimental evidence on human learning from predictive AI in cancer diagnosis**

Vivianna Fang He
UCL School of Management
University College London
One Canada Square, London E14 5AA, United Kingdom
Vivianna.he@ucl.ac.uk

Sihan Li
INSEAD
One Ayer Rajah Avenue, SG 138676, Singapore
sihan.li@insead.edu

Phanish Puranam
INSEAD
One Ayer Rajah Avenue, SG 138676, Singapore
phanish.puranam@insead.edu

Feng Lin
West China Hospital
Sichuan University
Wangjiang Campus, No.24 South Section 1, Yihuan Road, Chengdu, China
linfeng0220@scu.edu.cn

**Acknowledgements:** The first three authors contributed equally. The authors thank Manav Raj (University of Pennsylvania), Whitney Zhang (Massachusetts Institute of Technology), Xinyu Liang (INSEAD), Vivek Choudhary (Nanyang Technological University), and Bin Guo (Zhejiang University) as well as seminar participants at CEIBS, ESADE, HKUST, King's College London; London Business School, NUS, and Peking University for helpful feedback on previous drafts.

**Tool and Trainer? Experimental Evidence on Human Learning from Predictive AI in Cancer Diagnosis**

**Abstract:** Artificial intelligence (AI) is increasingly deployed to aid professional judgment, yet its implications for the development of human expertise remain unclear. We examine how the availability of AI assistance during training and task execution shapes novices' individual and collective judgment in the context of cancer diagnosis. Across two field experiments at a large teaching hospital, 576 medical novices learned to diagnose lung cancer from CT scans. Study 1 used a 2 × 2 factorial design that randomized AI assistance during training, task execution, both, or neither. Novices who received AI assistance during training achieved higher diagnostic accuracy in both AI-assisted and unaided task execution, with a larger improvement in the former. However, training-phase assistance was associated with more common errors than task-execution-phase assistance, suggesting that while the trainer effect improves individual accuracy, it limits the benefits of aggregating judgments. Study 2 shows that AI assistance during earlier task execution improved subsequent unaided performance and reduced common errors in the unaided phase of task execution, thus revealing a surprisingly beneficial "tool-as-trainer" effect. These findings suggest that, contingent on the right deployment strategy, AI assistance can support individual expertise development and improve collective judgment by reducing common errors across individuals.

## 1. Introduction

Artificial intelligence (AI) is increasingly embedded in work processes to help professionals make decisions more quickly and accurately (Jia et al. 2024, Kerstan et al. 2025, Krakowski et al. 2026). Yet, while AI use can improve professional performance (e.g., Noy and Zhang 2023), it may also undermine the preservation and development of expertise (Krakowski et al. 2023, Kosmyna et al. 2025). This tension is particularly consequential for novices. Less-experienced and lower-skilled professionals could realize larger immediate performance gains from AI assistance (Caplin et al. 2026), but they also have more to lose from forgoing opportunities to develop expertise through learning by doing (Brynjolfsson et al. 2025, Dell'Acqua et al. 2026). Given the cost of undeveloped skills emerges later or accrues largely to workers rather than to managers making adoption decisions (Puranam 2024), firms facing competitive pressure often prioritize short-term productivity while overlooking distant goals (Levinthal and March 1993). AI use therefore creates a paradox: The technology can improve aided performance in the short term while eroding human expertise development and the talent pipeline, which remain valuable for complementing and evaluating AI systems.

Existing evidence on how predictive and generative AI affect human expertise is inconclusive. A set of studies on strategic games suggests that AI can support human learning and task performance (e.g., Gaessler and Piezunka 2023, Choi et al. 2025, Bastani et al. 2026). In contrast, recent evidence shows that AI improves task performance only while assistance is available and leaves users less capable when they must work on their own (e.g., Kosmyna et al. 2025, Liu et al. 2026; Shen and Tamkin 2026). These divergent findings suggest that AI can operate either as a substitute for human learning or as a scaffold that supports it (Autor et al. 2026), but it remains unclear what determines which outcome arises.

Emerging research suggests that AI deployment strategies may partly explain the divergent evidence. Deployment strategies differ along two key dimensions, the type of AI assistance provided and the timing of that assistance. Across contexts, researchers found that declines in independent performance were concentrated among participants who received complete solutions from AI and thus were able to bypass cognitive engagement with the task at hand; those that received partial answers or hints suffered less (e.g.,

Bastani et al. 2025, Liu et al. 2026). In terms of timing, prior work (e.g., Kosmyna et al. 2025, Yin et al. 2025) suggests that AI assistance available before individuals (try to) understand the task impedes the cognitive processes that build or refine mental models, thereby undermining expertise development (Chi et al. 1981, Ericsson and Charness 1994). AI assistance with ready-to-copy solutions, especially when delivered before the user has engaged with the problem, appears to discourage users from developing their own capacity to solve the problem. Put differently, this type of deployment deprives users of opportunities for experiential learning (e.g., Argote 2012; Argote et al. 2021).

However, instead of deployment as a tool that substitutes for human cognition, another possibility, so far unexplored, is to deploy AI explicitly as a trainer that helps individuals learn vicariously. Vicarious learning, or learning from the experience of others, has a rich tradition in management science (Cyert and March 1963, Bandura 1977, Argote and Ingram 2000), yet existing studies have focused largely on the role of AI either as a tool or as a teammate for task completion (e.g., Anthony et al. 2023, Brynjolfsson et al. 2025, Kerstan et al. 2025), and yet to examine AI's trainer role systematically. Moreover, if AI shapes the mental models formed by human users, deployment strategy could affect not only individual accuracy but also collective accuracy. This is because the deployment of AI will shape the homogeneity of human errors, which in turn influences the decision quality of human groups. Both individual and collective accuracies matter in settings where judgments are aggregated, as in investments, recruitment, and diagnostic committees. A better understanding of these two critical trade-offs, between immediate performance and expertise development and between individual and collective performance, is important for the theory and practice of AI deployment in professional work.

Through two preregistered field experiments at a large teaching hospital in China, we study whether and how predictive AI assistance helps medical novices develop diagnostic expertise. Study 1 used a 2 × 2 factorial design that independently manipulated AI assistance during training and task execution. Participants received no AI assistance, AI assistance only during training, AI assistance only during task execution, or AI assistance during both phases. This design allows us to compare the effect of training-phase AI assistance (the "trainer" effect) on AI-assisted and unaided task performance. Study 2 examines whether a trainer effect can arise from AI assistance during task execution (the “tool-as-trainer” effect) and whether such an effect could persist with AI input reduced to malignancy probabilities alone, without highlighting scan features related to the prediction (the “taciturn trainer” effect). Across both studies, we evaluate diagnostic accuracy, precision, recall, and confidence-weighted accuracy (CWA). We also examine how each deployment strategy shaped the convergence of mental models by measuring common errors across individuals within each condition.

Our research sheds light on the role of AI in professional development, a topic with vast societal impact. Rather than treating AI primarily as a tool that improves assisted performance at the expense of human expertise, we examine whether it can act as a trainer when deployed across different phases of novices’ expertise development. Our results have implications for both theory and practice. First, in demonstrating that AI assistance improves medical novices’ diagnostic accuracy when it is made available during training and task execution, we show that AI can act as both a trainer and a tool. Our finding that AI assistance in the earlier task-execution phase produces improvements in subsequent unaided performance also contributes to resolving a long-standing debate in AI use around the trade-off between performance gain and expertise development (e.g., Gaessler and Piezunka 2023, Yin et al. 2025, Bastani et al. 2026). Second, our findings on how AI assistance in different phases facilitates the development of task-relevant mental models provide empirical evidence that AI assistance could enable different types of vicarious learning (through observation versus interaction, Almeida and Kogut 1999, Baum et al. 2000, Posen and Chen 2013). Third, we show that individual learning gains do not translate uniformly into collective gains, extending the classic learning dilemma (March 1991) into a context where humans learn from not just fellow organizational members but also (predictive) AI models. Finally, in revealing that AI assistance in different phases of professional development shapes how accurately novices judge and how independently they err, we inform the design of AI deployment strategy in settings where individual judgments are aggregated.

## 2. Theoretical Background and Expectations

To lay a foundation for our argument, we first review prior work and the mixed findings on the AI's impact on skill development. We then invoke the difference between experiential and vicarious learning to propose the possibility of a trainer effect beyond the better-known tool effect in AI use. We conclude this section by summarizing our theoretical expectations.

### 2.1. AI-Assisted Performance versus Human Expertise Formation

A large body of research has examined whether access to AI at the time of use improves the outcome produced by a human (e.g., Gaube et al. 2023, Brynjolfsson et al. 2025, Yin et al. 2025, Dell'Acqua et al. 2026). Researchers examine a set of outcomes that we refer to as AI-assisted performance, which reflects the level of performance produced while AI assistance is available. This performance depends on the user's capabilities, the information supplied by the AI, and the user's ability to incorporate that information. By contrast, performance unaided by AI (after prior access to AI) reflects what the user can do independently. Unaided performance thus indicates whether knowledge acquired through AI-assisted task completion has become independently deployable human expertise, in other words, whether the user has learned from using AI.

This distinction is particularly consequential for novices. Because novices possess less prior domain expertise, AI has great potential to improve their assisted performance (Caplin et al. 2026). However, using AI may impose a large risk of impeding the independent effort necessary for forming task-relevant mental models that constitute expertise (Chi et al. 1981, Ericsson et al. 1993. Therefore, the individuals who could benefit most from AI at the point of deployment (i.e. novices) may also have most to lose.

### 2.2. Mental Model Formation through Experiential Learning

The body of work on experiential versus vicarious learning is instrumental to understanding why AI assistance may support or undermine novices' expertise formation. Experiential learning, be it for individuals or for organizations, occurs when performing a task changes the knowledge subsequently available to similar tasks (Kolb 1984, Gavetti and Levinthal 2000, Argote 2012, Argote et al. 2021). This knowledge, gained from first-hand experience, is manifested in mental models, which are organized representations of perceptual or abstract objects and the relationships among them (Brandon and Hollingshead 2004; Waller et al. 2004). Experience alone is not sufficient to form such models. Learners must attend to relevant evidence, generate or retrieve a decision rule, act on that rule, observe feedback, and revise their understanding of the relationship between actions and outcomes (Argote and Miron-Spektor 2011). Errors can be particularly informative (Lei and Naveh 2023): When learners discover that an answer is incorrect, the discrepancy helps reveal the limits of the rule that produced it. Independent attempts, including failed ones, can therefore create learning opportunities that successful task completion with external support does not necessarily provide (He et al. 2018, Edmondson 2023).

For instance, in domains like medical-image diagnosis, the cognitive operations through which mental models form in experiential learning include searching the image, recognizing potentially diagnostic features, distinguishing relevant from irrelevant variation, integrating multiple features, translating the resulting representation into a probability or classification, and calibrating confidence in that classification (Lebovitz et al. 2022). Repeating these operations allows a novice to construct and refine the feature-output (i.e. diagnosis) relationships that constitute diagnostic expertise (Yin et al. 2025).

AI deployment can disrupt experiential learning by supplying complete solutions for users to adopt, without their having to map task features to outcomes themselves. Users may economize on cognitive effort when ready-to-use outputs are available. Additionally, AI systems being used may also reduce the observability of task-relevant information, not just the need to process it. For instance, a diagnostic AI system that provides a clinical decision without showing the patient the characteristics that led to the decision likely deprives users of opportunities for experiential learning (Sundaresan and Guler 2025). The same problem arises when a coding tool generates an entire code base from a one-line prompt but leaves the immediate next step (specifying a code architecture used to build the specific target functions) invisible (Shen and Tamkin 2026). In both cases, users lack access to important task features needed to construct their own input-output mapping, as they observe at best a prompt-to-output mapping.

Accumulated empirical evidence of impaired unaided performance following AI use is consistent with the theoretical reasoning that AI disrupts experiential learning. After using an unmodified version of GPT that provided complete solutions to mathematics problems, students performed worse when GPT access was removed (Bastani et al. 2025). Developers who used AI to learn an unfamiliar programming library later showed weaker conceptual understanding, code-reading ability, and debugging skill, particularly when they delegated code generation or debugging to the AI (Shen and Tamkin 2026). Participants who used AI for reasoning and comprehension tasks subsequently performed worse and were less willing to persist after assistance was withdrawn, with declines concentrated among those who obtained direct solutions (Liu et al. 2026). Kosmyna et al. (2025) report a similar pattern in essay writing: Extensive LLM reliance was associated with weaker memory encoding and neural connection, as well as reduced independent performance. At the organizational level, Sundaresan and Guler (2025) find that adoption of clinical recommendation systems was associated with flatter experiential learning curves, consistent with algorithmic reliance interfering with knowledge formation.

The common feature across the above findings, which cover both generative and predictive AI systems, is an AI deployment strategy that substitutes human cognition. With such deployment, users obtained outputs they could adopt without thinking about the underlying feature-outcome relationship (i.e. the relevant mental model). These findings therefore reveal a potential risk associated with using AI assistance as a substitute for building one's own mental models through experiential learning.

### 2.3. Mental Model Formation through Vicarious Learning

A mental model of the task landscape both reflects what's learnt and facilitates further learning (Baumann et al. 2019, Knudsen and Srikanth 2014). We propose that the same AI capabilities that would impede mental model formation can also support its formation when deployed differently. Rather than supplying complete solutions that allow users to bypass input-output mapping, AI can make this mapping more observable, thus enabling vicarious learning.

In contrast to experiential learning, vicarious learning provides individuals (and organizations) a way to learn without incurring the potential cost of failure (Bandura 1977, Argote and Ingram 2000; Denrell 2003, Denrell et al. 2026). Depending on the richness of the interaction that allows the individual to benefit from the other's experience, vicarious learning can be further categorized into contact learning (e.g., Almeida and Kogut 1999) that occurs through in-depth interactions that allow sharing mental models or beliefs that are otherwise difficult to transfer (Szelanski 1996, Bechky 2003) or mimetic learning (e.g., Baum et al. 2000, Posen and Chen 2013) that occurs through observation without direct interaction. Further, mimetic learning could involve different forms of information. A complete observation would include input-output mapping (or the relationship between the actions and their corresponding outcomes) and the explanations (or, beliefs) for that mapping. Yet, even without the explanations, as long as there is a substantial common ground of shared concepts and terminology, the learner may be able to infer the beliefs underlying the mapping (Clough and Piezunka 2020, Park and Puranam 2024).

In medical-image diagnosis, for example, analyzing historical patient cases or reading medical images annotated by digital decision support tools provides a source of vicarious learning (Gaube et al. 2023). In the general framework of machine and human learning, making the input-output relationship salient is equivalent to providing labeled data from which a learner (an algorithm or a human novice) can reverse engineer their own model or identify the category to which a situation belongs (Sutton and Barto 1998).

Access to (correctly) labeled data should be particularly useful during novice training, when mental models are still being developed (Gaube et al. 2023). For a novice confronting an unfamiliar task domain, their limited experience means the features relevant to the outcome are not self-evident. Unguided early practice therefore generates few instructive examples, because the novice cannot reliably identify what merits attention. Put differently, novices face a bootstrapping problem, because learning from experience requires some prior understanding of which features matter. AI assistance offers a solution to this problem. By supplying cases in which task-relevant features and diagnostic outcomes are both visible, AI can support vicarious learning through exemplars of input-output mapping (Baum et al. 2000, Posen and Chen 2013). Such exemplars then help novices develop an initial representation of which features matter and how they combine, without requiring extensive learning by doing or expertise they have yet to acquire (Darr et al. 1995, Argote 2012). Observational data is especially valuable when labels are largely correct and more representative than direct experience, as it can reduce noise and bias in the learning process (Denrell 2003). For humans as for machines, vicarious learning can complement the more costly and time-consuming process of experiential learning through trial and error (Sutton and Barto 1998).

There is broad evidence that AI systems can support vicarious learning. For example, gamers who were given AI advice improved subsequent decisions beyond merely following the advice (Bastani et al. 2026). Professional Go players improved their unaided tournament decisions after observing the moves of an AI that outperformed humans (Choi et al. 2025), and chess computers expanded access to experiential practice by serving as scalable training partners (Gaessler and Piezunka 2023). In this set of studies on strategic games, however, AI was available across extended periods spanning multiple phases of work, including structured study, tournament preparation, and post hoc analysis. Therefore, it is difficult to attribute human performance improvement to a specific phase of exposure to AI assistance.

Evidence from medical diagnosis provides additional support for a potential vicarious learning effect. AI-generated feature annotations improve contemporaneous diagnostic performance among non-specialist physicians (Gaube et al. 2023), and professionals integrate AI predictions more effectively when they can relate them to interpretable evidence (Lebovitz et al. 2022). Yet, neither study examines whether training-phase AI assistance produces mental models that persist after the AI is withdrawn. Prior evidence therefore is suggestive but does not identify training-phase AI assistance as a mechanism that builds independently deployable competence through vicarious learning. We turn next to developing and specifying our theoretical expectations of how AI input availability during training and task execution should shape novices' aided task performance and independent expertise.

### 2.4. Expected Effects of AI Assistance

Skill acquisition proceeds through phases that contribute differently to expertise (Ericsson 2018). Training introduces the learner to the structure of the professional domain and helps construct an initial mental representation of problems in that domain. In task execution or practice, the learner retrieves and applies that representation in new cases, discriminate between similar alternatives, experience uncertainty, and as a result may revise that mental representation through feedback (Csaszar and Levinthal 2016). Thus, AI assistance may have different effects depending on both the *phase* of the developmental trajectory (i.e., *when* it is provided) and the *content* of the observational data that it provides (i.e., *how* it is provided).

#### 2.4.1. Baseline: The Tool Effect

Novices enter diagnostic tasks with less developed mental models than experts and may therefore struggle to identify and integrate relevant features (Chi et al. 1981, Ericsson et al. 1993). Predictive AI can supplement these models by supplying case-specific predictions and directing attention to diagnostically relevant information. When AI predictions are more accurate than novices' own judgments and are appropriately incorporated into their decisions, AI assistance should improve performance on the current task, thereby producing a tool effect. This expectation is consistent with the evidence presented in previous studies on AI-assisted performance in various domains (e.g., Gaube et al. 2023, Brynjolfsson et al. 2025). We would, however, note that a tool effect does not require learning to have taken place. A novice can reach a more accurate diagnosis than they themselves could by using an AI tool without acquiring the ability to diagnose new cases independently. We therefore treat the tool effect as a baseline prediction about the performance when using an AI system with accuracy superior to the user.

### 2.4.2. The Trainer Effect

Rather than discovering each relationship through their own successes and failures, individuals can learn from exemplars of input-output mapping. Learning from experience requires some understanding of which features merit attention, yet acquiring that understanding itself requires learning. AI assistance in training helps the learner (especially novices) overcome this problem by making task-relevant features and associated predictions jointly observable without the learner first having to acquire much experience. The core mechanism here is vicarious learning through observation rather than experiential learning through independent task completion (Bandura 1977, Baum et al. 2000, Posen and Chen 2013). Observing AI-annotated exemplars helps the learner identify relevant features and construct an initial mental representation of how those features relate to diagnostic outcomes. Across cases, this mental representation becomes applicable beyond the particular examples encountered during training.

The expected training benefit of AI-provided exemplars depends on their information being suitable for learning. Relevant features must be observable, and the associated predictions must be sufficiently reliable and informative to guide model formation. Under these conditions, AI assistance supplies the input-output mapping from which the learner can build a mental model. The distinctive implication of the trainer effect is that its benefit should remain observable when AI is no longer available. We therefore expect that *among novices performing diagnostic tasks without AI assistance, those who received AI assistance during training achieve higher diagnostic performance than those who received no AI assistance during training.*

### 2.4.3. The Combined Tool and Trainer Effect

Providing AI assistance in both phases (training and task execution) combines a knowledge endowment acquired through observation (i.e., the initial mental model) with additional learning opportunities made available via contact (i.e., the refinement or revision of that mental model). Relative to AI assistance in the task execution only, combined deployment gives the learner an opportunity to construct that model before encountering the actual task. Relative to AI assistance in the training phase only, combined deployment adds case-specific AI predictions that can augment or challenge the learner's newly formed mental model. These comparisons suggest an advantage for combined deployment over assistance in either phase alone.

Training may also improve the basis for utilizing subsequent AI predictions. In the medical context, a physician with a more developed mental representation of diagnostic features can relate a prediction to task-relevant evidence rather than treating it as an isolated diagnosis. This reasoning is consistent with the importance of explanations of AI diagnosis for professionals' integration of AI predictions (Lebovitz et al. 2022). Training and task-execution AI assistance address different limitations, in that AI assistance in training supports the formation of an initial representation of the task domain, whereas that in task-execution supplies information about the current case to refine or revise that representation. Further, the first mechanism reinforces the second: A learner who enters task execution with a more accurate initial model is better positioned to evaluate each AI prediction. We therefore expect that *providing AI assistance in both phases (as a trainer during training and as a tool during execution) to improve the learner's performance beyond AI assistance in training alone.*

## 3. Experimental Design

We studied the effects of predictive AI assistance on novices' performance and expertise development in two preregistered field experiments in the context of lung cancer diagnosis.[1] Study 1 distinguished the tool, trainer, and combined effects. Its results led us to explore further variations of AI deployment and their learning effects in Study 2, in which we investigated whether a trainer effect could arise from tool use alone (i.e., a tool-as-trainer effect) and whether training effects persisted when training-phase assistance provided less explanatory information (i.e., a taciturn-trainer effect).

[1] Study 1: https://aspredicted.org/xbjc-ty6y.pdf and Study 2: https://aspredicted.org/psc4-2hnj.pdf.

### 3.1. Sample and Procedure

We conducted the experiments at a large teaching hospital in China. The protocol was approved by two independent institutional review boards and all participants provided informed consent. Study 1 included 336 medical trainees who were novices in radiological diagnosis (mean age = 22.783 years; 48.810% female). Study 2 included a nonoverlapping sample of 240 novice medical trainees (mean age = 22.629 years; 50% female). Participation was voluntary. To encourage engagement, the top 3% performers received awards of RMB 200 (approximately US$29), roughly 40% of the national minimum monthly stipend for Master's students in China.

In both studies, participants first attended a one-hour lecture covering conceptual knowledge about lung disease and CT-based diagnosis. They were then randomly assigned to a 25-minute workshop covering 10 real cases, during which they either did or did not receive AI assistance. All workshops used videos produced by the same instructor and the same case-based format. In the AI-assisted workshop, each case included a bright-green contour identifying the nodule, a malignancy-probability estimate, and seven quantified diagnostic features (CT attenuation value, scan layer, nodule location, volume, category, composition, and dimensions). In the no-AI workshop, the instructor covered the same cases without AI-generated annotations or predictions. Participants then proceeded to task execution, during which they classified 30 high-resolution chest CT scans as benign or malignant and rated their confidence in each judgment on a 1-10 scale. Depending on the condition, AI assistance was provided throughout, withheld throughout, or withdrawn partway through this phase.

Study 1 used a 2 × 2 factorial design (Figure S1) that independently randomized the availability of predictive AI assistance during training for the task and during subsequent task execution. Participants were randomly assigned to the four experimental conditions ($n$ = 84 per condition). Group A1 received no AI assistance in either phase, Group A2 received AI assistance during task execution only, Group B1 received AI assistance during training only, and Group B2 received AI assistance in both phases. This design allows us to identify the tool effect by comparing A2 with A1, the trainer effect by comparing B1 with A1, and the combined effect by comparing B2 with other single-phase conditions.

Study 2 used a reduced factorial design (Figure S2). Group 1 was equivalent to the no-AI control used in Study 1: Participants received no AI assistance during training and diagnosed all 30 task cases unaided. Group 2 received no AI assistance during training, had full AI assistance during cases 1-15 of task execution, and no assistance during cases 16-30. This was meant to test a tool-as-trainer effect, which seemed plausible given our Study 1 results. Group 3 received reduced AI input during training and no AI during any task-execution case. The reduced AI input displayed identified nodules and its malignancy probability but withheld the six quantified diagnostic features that were available in Study 1 (Figure S3). This reduction in the information was to test if our trainer effect in Study 1 could be explained as a transfer of a ready-made mental model from AI to novices rather than an assisted mental model formation by the novice.

### 3.2. Predictive AI System, Cases, and Ground Truth

The AI assistance was generated by a predictive system developed by LinkDoc, an oncology-focused healthcare big data and AI technology company. The system was trained on more than 40,000 domestic and international patient cases with histopathological ground truth. It combined a Faster R-CNN region-proposal network with 3D-ResNet18 for nodule detection and a three-dimensional convolutional neural network for malignancy classification. A six-member clinical faculty panel—two senior, two mid-career, and two junior faculty members—selected the 30 task cases to span three difficulty levels (10 easy, 10 medium, and 10 hard). The set contained 15 malignant and 15 benign cases. Using a 0.500 probability threshold, the AI correctly classified 25 of the 30 cases (83.333% accuracy); its area under the receiver-operating-characteristic curve was 0.920 (95% bootstrap CI [0.804, 0.991]; Figure S4). Participants who received AI assistance were explicitly informed that AI predictions were not always correct and that they remained responsible for making the final diagnostic judgment.

Rather than relying on expert consensus, the ground truth of our 30 task cases was established by following the most rigorous standard in medical research. Malignant nodules were confirmed by histopathological examination following surgery. Benign nodules were confirmed by the absence of clinically significant change over three years of follow-up. These criteria provided a common gold standard for evaluating both participant and AI judgments.

### 3.3. Dependent Variables

#### 3.3.1. Diagnostic Performance

We calculated four outcomes of individual-level diagnostic performance, including accuracy, precision, recall, and CWA. Accuracy was the proportion of correct classifications, precision was the number of true-positive diagnoses divided by all positive diagnoses, recall (sensitivity) was the number of true-positive diagnoses divided by all malignant cases, and CWA incorporated correctness and expressed confidence. For each case, the CWA contribution equaled +1 for a correct diagnosis and −1 for an incorrect diagnosis, multiplied by confidence on a 1 to10 scale. We summed these contributions across all 30 cases in Study 1 and cases 16-30 in Study 2. CWA could range from −300 to +300 in Study 1 and from −150 to +150 in Study 2, with higher values indicating more correct and more confidently correct judgments.

#### 3.3.2. Mental Model Development

We used two proxies for mental model development: participants' ability to distinguish correct from incorrect AI recommendations and the calibration of their confidence to their diagnostic accuracy.

To assess participants' ability to distinguish correct from incorrect AI recommendations, we coded, at the participant-case level, whether a participant's binary judgment diverged from the AI classification at the standard 0.500 probability threshold. As a sensitivity analysis, we reclassified AI recommendations using a 0.950 (high confidence) malignancy-probability cutoff and repeated all analyses. In Study 1, we compared divergence between A2 and B2 separately for AI-correct and AI-incorrect cases. In Study 2, where only Group 2 received AI assistance during part of task execution, we compared each participant's divergence rate on AI-correct and AI-incorrect cases using the 0.500 threshold.

Following Caplin et al. (2025), we defined participant $i$'s calibration as the negative absolute gap between mean stated confidence and actual accuracy across the cases diagnosed by that participant. Because confidence in the selected option of a binary classification lies between 0.500 and 1.000, we linearly transformed each raw confidence rating on the 1 to 10 response scale as $K_{ij} = 0.50 + (confidence_{ij} - 1)/18$, so that a rating of 1 maps to 0.500 and a rating of 10 maps to 1.000. Let $K_i$ denote participant $i$'s mean transformed confidence across the cases and let $p_i$ denote mean accuracy across those same cases. Calibration was defined as:

$$Calibration_i = -|K_i - p_i|$$

A value of 0 indicates perfect calibration; more negative values indicate a larger mismatch between mean confidence and actual accuracy. We computed the measure for all 30 task cases in Study 1 and for the unaided cases 16-30 in Study 2. Each participant thus received a single calibration score.

### 3.3.3. Common Errors

To measure errors common among participants within a condition, we constructed a common error index (E). For each participant, we identified every misclassified scan and counted how many other participants in the same condition also erred on that scan. We then averaged these counts across the focal participant's errors. For example, suppose participant $j$ misclassified scans 2 and 4. If five other participants erred on scan 2 and six erred on scan 4, then $E_j = (5 + 6)/2 = 5.500$. We compared mean E across conditions after excluding participants with perfect accuracy, for whom the measure is undefined. Larger values indicate that a higher error homogeneity within the condition.

### 3.3.4. Dependence Penalty

To evaluate the implications of common errors for collective performance under a three-person majority-vote rule, we first enumerated all possible within-condition triads. For each triad and scan, the majority diagnosis determined the committee decision; comparison with ground truth yielded collective accuracy ($A_c$). We then assessed robustness to committee composition using 5,000 Monte Carlo replications. Each replication sampled participants without replacement to form 28 disjoint committees per Study 1 condition and 26 committees per Study 2 condition (leaving two participants unused). For each sampled committee, we applied majority voting to every analyzed scan; run-level Ac was the mean accuracy across sampled committees and scans.

For the same committees, we also computed an independence benchmark: the expected committee accuracy if the three members' errors were statistically independent. Letting $p_i$ denote participant $i$'s individual accuracy across scans and $q_i = 1 - p_i$ their individual error rate, the probability that a three-person majority vote returns the correct answer under independence is the probability that at least two of the three members are individually correct: $p_1 p_2 p_3 + p_1 p_2 q_3 + p_1 q_2 p_3 + q_1 p_2 p_3$. Averaging this value across all sampled committees yielded the independence benchmark $P_c$.

The dependence penalty is defined as $d = P_c - A_c$. A positive value of $d$ indicates that the committee underperformed the independence benchmark because participants' errors were positively correlated within the condition. The Monte Carlo analysis repeated the participant-triad sampling 5,000 times for each condition while evaluating every sampled committee on all analyzed scans, yielding the distributions of $d$.

## 3.4. Analytical Strategy

Randomization checks used analysis of variance for age and a Pearson chi-square test for gender. Individual-performance analyses used the participant as the unit of analysis. For accuracy, precision, recall, and CWA, we estimated one-way analyses of variance. Table 1 reports all pairwise mean comparisons using Tukey's HSD to control the familywise error rate within each outcome and study, together with the corresponding 95% confidence intervals and $p$ values. The B1-A2 recall comparison discussed with the common error results is a separate two-sided Welch contrast, as indicated there and in the Table 1 note. Calibration analyses used ordinary least squares with the factorial terms in Study 1 and condition indicators in Study 2. Companion performance models added age, gender, and prior AI diagnostic experience. Study 1 divergence rates were compared with two-sided equal-variance *t* tests at the participant-case level, separately for AI-correct and AI-incorrect cases. Study 2 used a two-sided paired *t* test of participant-level divergence rates on AI-incorrect and AI-correct cases. Common error comparisons used participant-level Welch tests. For the dependence penalty *d*, we used Welch tests for differences in means and Kolmogorov-Smirnov tests for differences in distributions.

-- Insert Table 1 about here --

# 4. Experimental Results

## 4.1. Study 1 Results

Random assignment produced closely balanced conditions. Mean age did not differ across conditions (all pairwise $p > 0.990$; Table S1). Each condition included 41 women and 43 men, yielding identical gender composition (Pearson $\chi^2(3, N = 336) = 0.000$, $p = 1.000$; Table S2).

### 4.1.1. Diagnostic Performance

Figure 1 shows that predictive AI assistance improved individual performance when deployed during task execution. Among participants trained without AI assistance, task-execution assistance increased mean accuracy from 0.678 in the no AI control group to 0.755 in A2 (mean difference = 0.077, s.e. = 0.012, 95% CI [0.046, 0.107], $p < 0.001$) and increased CWA by 60.012 points ($p < 0.001$). Training-phase assistance also improved subsequent unaided performance. Among participants who completed the task without AI assistance, mean accuracy increased from 0.678 in the no AI control group to 0.710 in B1 (mean difference = 0.032, s.e. = 0.012, 95% CI [0.002, 0.063], $p = 0.034$), and CWA increased by 20.500 points ($p < 0.001$). The A2-A1 comparison identifies the contemporaneous tool effect, whereas the B1-A1 comparison identifies a trainer effect that persisted after AI assistance was withdrawn.

--- Insert Figure 1 about here ---

The combined deployment produced the strongest individual performance (Table S3). Mean accuracy was 0.813 in B2, compared with 0.678 in the no AI control group, 0.755 in A2, and 0.710 in B1. B2 exceeded the no AI control group by 0.135 (s.e. = 0.012, 95% CI [0.104, 0.165]), A2 by 0.058 (95% CI [0.028, 0.089]), and B1 by 0.103 (95% CI [0.072, 0.133]); all $p < 0.001$. CWA followed the same ordering (A1 = 77.238, A2 = 137.250, B1 = 97.738, B2 = 159.583). Relative to the no AI control group, B2 increased CWA by 82.345 points (s.e. = 4.749, 95% CI [70.082, 94.608], $p < 0.001$).

The single-phase deployments improved accuracy through a different error trade-off from the combined deployment. Relative to the no AI control group, precision increased by 0.150 in A2 (s.e. = 0.014, 95% CI [0.113, 0.187], $p < 0.001$) and by 0.047 in B1 (95% CI [0.010, 0.084], $p = 0.006$). Recall decreased by 0.083 in A2 (s.e. = 0.019, 95% CI [−0.132, −0.033], $p < 0.001$). The 0.041 decrease in B1 was not statistically significant (95% CI [−0.090, 0.008], $p = 0.134$). Thus, execution-only assistance reduced false positives but increased false negatives. Training-only assistance produced the same directional pattern, although its recall difference was not statistically distinguishable from zero. In B2, precision exceeded the control group by 0.159 (s.e. = 0.014, 95% CI [0.122, 0.196], $p < 0.001$), whereas recall did not differ statistically from the control group (mean difference = 0.044, s.e. = 0.019, 95% CI [−0.006, 0.093], $p = 0.102$).

### 4.1.2. Evidence for Mental Model Development

Participants who received AI assistance during training more effectively distinguished correct from incorrect AI recommendations during task execution (Table S4). When the AI classification was correct at the standard 0.500 probability cutoff, mean divergence was 0.228 in A2 and 0.155 in B2 (B2 − A2 = −0.072, s.e. = 0.012, 95% CI [−0.096, −0.049], $t(4{,}198) = -5.985$, $p < 0.001$). This difference shows those who received AI assistance during training diverged significantly less from the correct recommendations than those who did not. When the AI classification was incorrect at this threshold, mean divergence did not differ between A2 and B2 (0.667 vs. 0.655; B2 − A2 = −0.012, s.e. = 0.033, 95% CI [−0.076, 0.052], $t(838) = -0.364$, $p = 0.716$). However, when using a 0.950 malignancy-probability cutoff for AI classification, on AI-incorrect cases participants who received AI assistance during training diverged from AI significantly more often than those who did not (0.812 vs. 0.705; B2 − A2 = 0.107, s.e. = 0.018, 95% CI [0.071, 0.143], $t(2,182) = 5.895$, $p < 0.001$). Taken together, these results suggest that training-phase assistance improved novices' ability to judge when to follow or override AI recommendations rather than producing uniform deference to AI.

Analysis using calibration as an outcome variable provided additional evidence of mental model development. Mean calibration, where values closer to zero indicate better alignment between confidence and accuracy, was −0.136 in the no AI control group A1, −0.076 in A2, −0.102 in B1, and −0.062 in B2. Task-execution AI assistance improved calibration ($\beta = 0.060$, s.e. = 0.008, 95% CI [0.043, 0.077], $p < 0.001$), as did training-phase AI assistance ($\beta = 0.034$, s.e. = 0.008, 95% CI [0.017, 0.050], $p < 0.001$). Their interaction was not statistically significant ($\beta = -0.020$, s.e. = 0.012, 95% CI [−0.043, 0.004], $p = 0.103$). This sets of results suggest that AI assistance in both deployment phases brought confidence into closer alignment with diagnostic accuracy, that is, better calibration. More specifically, participants who had AI assistance in training later showed better calibration during their independent task execution than the no AI control (the absolute difference between B1 and A1 group was 0.034, 95% CI [0.017, 0.050] p < 0.001), indicating a mental model improvement benefit associated with AI assistance in training. accuracy and the best calibration.

#### 4.1.3. Common Errors

AI assistance generally reduced the error homogeneity within a condition, but the magnitude of reduction depended on deployment phase (Figure S6). Mean common-error scores were 47.148 in the control Group A1, 25.698 in Group A2, 41.302 in Group B1, and 21.241 in Group B2 ($p < 0.001$ for all pairwise comparisons). Relative to the no AI control group, training-only assistance reduced the score by 5.846 points, whereas execution-only assistance reduced it by 21.450 points. The execution-phase reduction was therefore 15.604 points larger (95% CI [14.260, 16.949], $t(160.576) = 22.925$, $p < 0.001$).

For the targeted B1-A2 Welch contrast described in Section 3.4, B1 participants had modestly higher recall than A2 participants (0.747 vs. 0.706; mean difference = 0.041, 95% CI [0.003, 0.080], $t(146.819) = 2.110$, $p = 0.037$).

#### 4.1.4. Dependence Penalty

Table 2 summarizes the collective performance of the synthetic committees. Majority voting improved accuracy relative to individual judgments in every condition as expected, but collective accuracy remained below the independence benchmark. In the no AI control group A1, mean individual accuracy of 0.678 increased to collective accuracy $A_c = 0.705$, below $P_c = 0.756$ and yielding $d = 0.051$. In B1, individual accuracy of 0.710 increased to $A_c = 0.733$, but $P_c = 0.797$ and $d = 0.064$ indicated a slightly larger dependence penalty than in the control group. Conditions with task-execution assistance were closer to the independence benchmark: A2 increased from 0.755 to $A_c = 0.826$ ($P_c = 0.849$, $d = 0.023$), and B2 increased from 0.813 to $A_c = 0.888$ ($P_c = 0.908$, $d = 0.020$). The Monte Carlo distributions showed the same ordering. Mean $d$ was 0.051 for A1, 0.064 for B1, 0.023 for A2, and 0.020 for B2 (see Figure 2); every pairwise comparison from both the Welch and Kolmogorov-Smirnov tests had $p < 0.010$. Task-execution assistance was therefore associated with higher individual accuracy, more diverse residual errors, and a larger gain from aggregating individual diagnoses.

--- Insert Table 2 & Figure 2 about here ---

### 4.2. Study 2 Design and Results

Study 1's results indicated that AI assistance during the task execution phase alone appeared to reduce common errors and improve aggregate outcomes. This suggests possible benefits for mental model development from having AI assistance in task execution. When relevant features and associated predictions are jointly observable during execution, novices could compare the AI's interpretation of a case with their own, revisit features they initially overlooked, and reconsider how those features inform a diagnosis. AI thus provided exemplars of the input-output mapping as the leaner attempted to apply their own developing models, enabling a form of vicarious learning that is based on contact (Almeida and Kogut 1999). Relative to unaided practice, these exemplars could make visible relationships that novices may struggle to discover due to their limited experience. An AI prediction, however, is not itself verified outcome feedback, and disagreement with AI does not establish that the learner is wrong, although disagreement could motivate reflection. Therefore, in Study 2, we tested if novices receiving AI assistance during task execution would achieve higher diagnostic performance on subsequent new cases assessed without AI assistance than novices receiving no AI assistance during task execution (i.e., the tool-as-trainer effect).

An additional consideration for the design of Study 2 was to investigate whether the trainer effect entailed novices constructing their own mental model using AI labelled data, or imposing on them the AI system's representation of the diagnostic problems (i.e., the rationale for the mapping connecting inputs to outputs). Vicarious learning requires the observability of input-output mapping, with or without an explanation of why such mapping exits or how it is constructed. Mimetic learning can occur without direct transfer of the beliefs underlying an observed mapping, as long as the learner understands enough of the task features and classification categories to interpret the exemplars (Clough and Piezunka 2020, Park and Puranam 2024). In medical-image diagnosis, AI can display relevant features together with a predicted classification without providing a verbal or causal account linking them. Such assistance is a form of silent training, because it provides the key observation necessary for learning, in that the features and prediction remain jointly observable. This mechanism is a form of vicarious learning through observation without explicit belief transfer (Bandura 1977, Baum et al. 2000, Posen and Chen 2013) or reproducing the AI's internal decision process. In Study 2, we therefore also tested whether among novices performing diagnostic tasks without AI assistance, those who received training-phase AI assistance that displayed task-relevant features and associated predictions without explanations achieved higher diagnostic performance than those who received no AI assistance during training.

Like in Study 1, random assignment also produced closely balanced conditions in Study 2. Mean age did not differ across conditions (all $p \geq 0.651$; Table S5), and gender composition was also similar (Pearson $\chi^2(2, N = 240) = 0.300$, $p = 0.861$; Table S6).

### 4.2.1. Diagnostic Performance

Participants in Group 2, who received full AI assistance during cases 1-15 and then lost access to it, outperformed the no-AI control on the subsequent unaided cases across all four performance measures (Figure 3). Mean accuracy was 0.877 in Group 2 and 0.778 in Group 1 (mean difference = 0.099, s.e. = 0.017, 95% CI [0.060, 0.138], $p < 0.001$). Group 2 also exceeded Group 1 in precision by 0.078 (95% CI [0.042, 0.114], $p < 0.001$), recall by 0.103 (95% CI [0.052, 0.154], $p < 0.001$), and CWA by 23.291 points (s.e. = 3.223, 95% CI [15.690, 30.892], $p < 0.001$). These differences show that task-execution assistance improved subsequent unaided performance, supporting the tool-as-trainer effect.

Group 3, which received reduced AI assistance during training—nodule highlighting and a malignancy-probability estimate—also outperformed the no-AI control. Mean accuracy was 0.824 in Group 3 and 0.778 in Group 1 (mean difference = 0.047, s.e. = 0.017, 95% CI [0.007, 0.086], $p = 0.015$), and CWA was 12.841 points higher (95% CI [5.240, 20.442], $p < 0.001$). Precision and recall were similarly higher, although evidence for these differences was not conclusive (precision: mean difference = 0.036, 95% CI [−0.000, 0.072], $p = 0.050$; recall: mean difference = 0.046, 95% CI [−0.005, 0.097], $p = 0.089$). These results suggest that AI assistance need not include explicit diagnostic features to function as a trainer; exposure to AI-labeled exemplars was sufficient to improve subsequent unaided performance.

--- Insert Figure 3 about here ---

### 4.2.2. Evidence for Mental Model Development

Group 2 participants' diagnoses in AI-assisted cases 1-15 show that they evaluated AI recommendations rather than following them uniformly. Mean divergence was 0.756 when the AI classification was wrong and 0.138 when it was correct. The within-participant difference was 0.618 (s.e.

= 0.032, 95% CI [0.554, 0.682], $t(79) = 19.245$, $p < 0.001$). This ability to accept correct recommendations while rejecting incorrect ones suggests that novices developed a mental model of the diagnostic task while using the AI.

Calibration on the unaided cases 16-30 provides evidence that learning during AI use persisted after assistance was withdrawn (Figure S5). Mean calibration was −0.093 in Group 1, −0.066 in Group 2, and −0.096 in Group 3. Participants who had AI in the earlier practice phase (Group 2) was better calibrated than Group 1(the no AI control) in the subsequent, unaided phase of practice (mean difference = 0.027, s.e. = 0.010, 95% CI [0.007, 0.047], $p = 0.010$). This improvement, together with participants' selective acceptance of AI assistance during cases 1-15, shows evidence of mental model development and helps explain the tool-as-trainer effect. Group 3 did not differ from Group 1 (mean difference = −0.003, s.e. = 0.010, 95% CI [−0.023, 0.017], $p = 0.783$), indicating that reduced-information training assistance did not seem to improve calibration in this setting.

### 4.2.3. Common Errors and Dependence Penalty

Mean common-error scores on cases 16-30 were 31.899 in Group 1, 18.357 in Group 2, and 29.417 in Group 3 (Figure S6). Relative to Group 1, AI assistance during the first half of task execution (Group 2) reduced the score by 13.542 points (95% CI [11.193, 15.890], $p < 0.001$), whereas reduced training-phase assistance (Group 3) lowered it by 2.481 points (95% CI [0.190, 4.773], $p = 0.034$). The difference between these reductions was 11.060 points (95% CI [8.942, 13.179], $t(132.815) = 10.325$, $p < 0.001$). Early task-execution assistance therefore reduced common errors substantially more than reduced training-phase assistance.

Table 2 shows that the committee's diagnosis improved upon individual diagnostic accuracy in every condition on unaided cases 16-30. Accuracy increased from the mean individual score of 0.778 to collective accuracy $A_c = 0.821$ in Group 1, from 0.877 to $A_c = 0.932$ in Group 2, and from 0.824 to $A_c = 0.861$ in Group 3. Nevertheless, collective accuracy remained below the Condorcet independence benchmark in every condition: $P_c = 0.874$ in Group 1, 0.958 in Group 2, and 0.919 in Group 3. The corresponding dependence penalties were $d = 0.052$, 0.026, and 0.058. Group 2 had the smallest dependence penalty after receiving AI assistance on cases 1-15 and completing cases 16-30 without it. By contrast, Group 3's dependence penalty was slightly larger than that of the no-AI control despite its higher individual accuracy. These results reveal a trade-off between improving average individual performance and preserving the diversity of errors needed for high collective accuracy.

The Monte Carlo distributions were consistent with the all-triad comparison: Mean $d$ was 0.053 in Group 1, 0.026 in Group 2, and 0.058 in Group 3. Group 2 had a lower mean $d$ than both Group 1 and Group 3, whereas Group 3 had a higher mean than Group 1. All pairwise Welch and Kolmogorov-Smirnov comparisons were significant at $p < 0.010$ (Figure 2). Thus, AI assistance in the early execution phase produced gains that persisted after withdrawal while also preserving more heterogeneous residual errors, allowing the higher individual accuracy to translate into more accurate collective judgment.

## 5. Discussion of Results

Through two preregistered field experiments in a teaching hospital, we investigate the conditions under which a predictive AI tool could function as a trainer. Study 1 shows that the AI system we deployed is a competent tool: Novices who had AI assistance available during task execution alone (Group A2) achieved substantially higher diagnostic accuracy than novices working unaided (Group A1). More importantly, the AI system functions as a trainer by acting as a channel for vicarious learning: Novices who received AI input only during training (Group B1) achieved higher unaided diagnostic accuracy and better calibration than novices trained without AI (Group A1). Deploying AI across both phases (Group B2) produced the highest accuracy, approaching the range observed among experienced radiologists, and improved both precision and recall. Such improvement was not obtained under either type of single-phase deployment, both of which improved precision but reduced recall.

To further explain why such trainer effect exists, we examine novices mental model improvement. Comparison of compliance with AI recommendations across groups, we find that AI assistance in the training phase is helps novices critically evaluate the AI recommendations, and as a result, diverge less when AI is correct. Comparison of calibration shows that AI assistance in either the training or task execution phase helps novices better understand their diagnosis, but when combined, it benefits calibration the most. Together, these two sets of results suggest that AI assistance helps novices build a stronger initial mental model, which enables them to evaluate the AI's predictions rather than defer to them uniformly and to assign correct confidence level to their own diagnoses. Such mental modal development underlies the trainer effect we observed in Study 1.

Study 1 also produced two findings we did not anticipate: First, AI assistance in the execution phase seems to be as beneficial for improving individual calibration as that in the task execution phase. Second, AI assistance in this phase reduced common errors more than assistance in the training phase, which seems to guide how novices approach cases uniformly. These results demonstrate that tool use could also build or tune individual mental models (e.g. by supplying case-specific inputs against which each novice can compare their own judgments). Furthermore, AI assistance in the task execution phase functions better in developing individual idiosyncratic mental models.

This relative benefit in developing individual idiosyncratic mental models relates to the fact that in Study 1 the AI assistance during training supplied a rich informational package. Study 1 AI assistance included the malignancy probability, a visual highlight of the identified nodule, and seven quantified diagnostic features (CT attenuation value, scan layer, nodule location, volume, category, composition, and dimensions). Could the trainer effect be explained away as the "AI essentially gave the novice a mental model"? Does the trainer effect reflect novice mental model construction or a transmission of a list of features that discriminate malignant from benign scans? Distinguishing these interpretations (either of which is a form of vicarious learning) required reducing the informational richness of the AI assistance during training and observing whether the trainer effect survived. We reasoned that if the trainer effect depended on the explicit feature description, removing it should eliminate the effect. If instead the trainer effect arose from repeated exposure to CT scans paired with expert-level probability estimates, the effect should persist even when the information was reduced to showing the nodule and its malignancy probability, without naming the features that drove the prediction.

The findings of Study 2 resolved the questions and extended the theoretical account we developed in Study 1. First, we found that AI assistance in the execution phase can produce learning that persists after the AI assistance is withdrawn. Novices in Group 2, who had AI available during only the first half of practice, outperformed the no-AI control on subsequent unaided cases across every diagnostic measure. This difference supports the tool-as-trainer effect and suggests that cognitive offloading is not an inevitable consequence of AI availability during task execution. When the deployment preserves observability of the input-output mapping and users are motivated to engage with it (as they were in our studies through strong incentives for accuracy), AI assistance that augments performance while present can also build capability that outlasts its removal. The tool can be a trainer even when it supplies complete answers, rather than withheld answers as deployed in the guarded tutor version of Bastani et al. (2025). What appears to matter is not necessarily the withholding of final outputs but the preservation of exposure to and the motivation to process the information containing the task relevant input-output relationships.

Second, we found that novices in Group 3, for whom the AI assistance in the training phase included only the nodule location and malignancy probability (without the diagnostic features that accompanied AI assistance in Study 1) still outperformed the no-AI control on subsequent unaided cases. The effect was smaller than the full-feature trainer effect in Study 1 and smaller than the tool-as-trainer effect in Study 2, suggesting that removing the AI explanations may weaken but not eliminate the trainer effect. This finding echoes previous studies on vicarious learning without transfer of beliefs (e.g., Clough and Piezunka 2020) and adds to the current debate on the value of explanatory AI in medical settings (e.g., Li et al. 2026). Importantly, this rules out the interpretation that the trainer effect observed in Study 1 was due to AI transferring an expert-authored mental model to the novice. Novices learn from seeing scans paired with expert-level probability estimates, much as a supervised learning algorithm estimates a model from labelled data without being handed the model itself. The trainer effect reflects novice mental model construction from labelled exemplars, not only mental model transmission from AI to human.

Third, we replicated the different impacts of AI assistance in the training versus execution phase observed in Study 1. In Study 2, Group 2, which had AI assistance in the first half of practice, had a much smaller number of common errors and produced committees significantly closer to the independence benchmark than the no-AI control, indicating that novices in this condition retained heterogeneous approaches after AI assistance was withdrawn. Group 3, which had reduced AI assistance in training, by contrast, reduced common errors less and produced committees no closer to the independence benchmark than the no-AI control, despite significantly improved individual accuracy. In sum, a deployment strategy that relies on AI assistance in the training phase could raise individual accuracy while homogenizing errors and would then reduce the informational value of second opinions. This nuance is consequential for professional settings where individual judgments are aggregated across colleagues or group members.

## 6. Concluding Remarks

Our studies show that whether AI deployment supports or undermines the development of novice expertise can be understood in terms of impediments to experiential learning versus boosting vicarious learning (e.g., Argote and Ingram 2000, Argote et al. 2021). We further demonstrate that deployment configurations best serving individual capability formation differ from those best preserving the diversity of errors from which collective judgment benefits. Before we discuss the implications of these findings, we must note that our findings should be interpreted within their boundary conditions. Unlike scenarios in which humans can copy and paste AI output for task completion, as when generative AI such as LLMs produced solutions, code, or written text (Bastani et al. 2025, Kosmyna et al. 2025, Liu et al. 2026, Shen and Tamkin 2026), in cancer diagnosis, the deployment of predictive AI leaves the task input observable and requires the human to process it to form a final diagnosis. In our empirical context, the CT scan remains fully visible to participants and the AI supplies its probability estimate alongside the scan. Participants engaging with this information are exposed to similar input-output mapping that supports vicarious learning during training, but now paired with their own attempted diagnosis on the same case. Moreover, our participants received strong performance incentives (monetary rewards for the top performers) which further motivated cognitive engagement. Such cognitive engagement is likely a prerequisite for AI assistance to function not only as a substitute for a user's mental model but also as a scaffold for developing it. In identifying the trainer effect of AI assistance and the boundary conditions of this effect, our study provides causal evidence that AI tools can help human learning.

Within these boundary conditions, our study provides new insights into two important trade-offs in AI deployment across two studies with different AI assistance configurations. First, at the individual level, AI assistance, when deployed independently in the training or task execution phase, reduced false positives (an outcome desirable for the hospitals from a resource efficiency perspective) but increased false negatives (a consequential outcome for cancer patients). AI assistance with the same content deployed in both phases, however, reduced both false positives and false negatives. Second, AI assistance, when deployed in the training phase, homogenizes how novices reason while improving their individual unaided judgment. In contrast, AI assistance deployed in the task execution phase improves individual judgment while preserving error diversity, revealing an underexplored relative advantage of vicarious learning through interaction over observation (e.g., Clough and Piezunka 2020). In exploring the group-level advantages associated with vicarious learning results, we complement emerging insights that reveal the benefit of having diverse learning targets (and thus varied mental models) for individual performance (Denrell et al. 2026).

More broadly, our results imply that evaluating AI deployment solely by immediate performance gain would miss consequential long-term issues. Prior studies document that AI deployment can impede human expertise formation despite of improving performance when the AI tool is present. Our results suggest that this may be due to a loss of information or motivation (or a combination of both) for experiential learning from engaging with raw cases. In addition, we find that AI deployment may increase the correlation of human errors within a group while improving individual performance, updating the old individual-organizational learning dilemma with new nuances (March 1991). We believe that these long-term issues are critical for the talent pipeline within organizations and for the collective intelligence that has propelled human societies, by shaping the development of individual expertise and preserving the value of aggregating individual expertise.

In professional domains where human involvement remains both necessary and desired, responsible AI deployment requires managers and policy makers to carefully evaluate the trade-offs revealed in our study. For domains that primarily rely on individual judgment, an AI deployment strategy that in the short term privileges reducing the most consequential errors, while in the longer term allowing for development of human expertise seems most appropriate. For domains that require a group process (e.g., juries, tumor diagnosis boards, syndicated investment, and academic hiring committees), an AI deployment strategy that preserves error diversity in the group is essential for high-quality group decisions.

Several limitations qualify these conclusions. We studied novices in one medical domain; effects may differ for experts (Allen and Choudhury 2022, Autor et al. 2026) or for other domains. While our results are consistent with preserved cognitive engagement as a driver of learning (Choi et al. 2025), the design does not measure the underlying cognitive or neurological processes. The experiments occurred over a limited time horizon with a bounded set of patient cases, leaving longer-term dynamics uncertain. Finally, we tested one predictive AI system in a diagnostic setting; generalization will require replication across AI systems, data modalities, and task domains, including generative AI settings in which the observability conditions we identify as consequential may differ. We hope that our research stimulates more future studies exploring the responsible deployment strategies of AI systems that benefit not just the performance today but also enable humans to continue to develop alongside these powerful technologies.

**Table 1. Pairwise comparisons of performance metrics (Study 1 and Study 2)**

| Metric | Comparison | Mean difference | s.e. | 95% CI | *p* value |
|---|---|---|---|---|---|
| Study 1 | | | | | |
| Accuracy | A2 vs. A1 | 0.077 | 0.012 | [0.046, 0.107] | < 0.001 |
| Accuracy | B1 vs. A1 | 0.032 | 0.012 | [0.002, 0.063] | 0.034 |
| Accuracy | B1 vs. A2 | −0.044 | 0.012 | [−0.075, −0.014] | 0.001 |
| Accuracy | B1 vs. B2 | −0.103 | 0.012 | [−0.133, −0.072] | < 0.001 |
| Accuracy | B2 vs. A1 | 0.135 | 0.012 | [0.104, 0.165] | < 0.001 |
| Accuracy | B2 vs. A2 | 0.058 | 0.012 | [0.028, 0.089] | < 0.001 |
| Precision | A2 vs. A1 | 0.150 | 0.014 | [0.113, 0.187] | < 0.001 |
| Precision | B1 vs. A1 | 0.047 | 0.014 | [0.010, 0.084] | 0.006 |
| Precision | B1 vs. A2 | −0.104 | 0.014 | [−0.141, −0.067] | < 0.001 |
| Precision | B1 vs. B2 | −0.113 | 0.014 | [−0.149, −0.076] | < 0.001 |
| Precision | B2 vs. A1 | 0.159 | 0.014 | [0.122, 0.196] | < 0.001 |
| Precision | B2 vs. A2 | 0.009 | 0.014 | [−0.028, 0.046] | 0.924 |
| Recall | A2 vs. A1 | −0.083 | 0.019 | [−0.132, −0.033] | < 0.001 |
| Recall | B1 vs. A1 | −0.041 | 0.019 | [−0.090, 0.008] | 0.134 |
| Recall | B1 vs. A2 | 0.041 | 0.019 | [−0.008, 0.090] | 0.134 |
| Recall | B1 vs. B2 | −0.085 | 0.019 | [−0.134, −0.036] | < 0.001 |
| Recall | B2 vs. A1 | 0.044 | 0.019 | [−0.006, 0.093] | 0.102 |
| Recall | B2 vs. A2 | 0.126 | 0.019 | [0.077, 0.175] | < 0.001 |
| CWA | A2 vs. A1 | 60.012 | 4.749 | [47.749, 72.275] | < 0.001 |
| CWA | B1 vs. A1 | 20.500 | 4.749 | [8.237, 32.763] | < 0.001 |
| CWA | B1 vs. A2 | −39.512 | 4.749 | [−51.775, −27.249] | < 0.001 |
| CWA | B1 vs. B2 | −61.845 | 4.749 | [−74.108, −49.582] | < 0.001 |
| CWA | B2 vs. A1 | 82.345 | 4.749 | [70.082, 94.608] | < 0.001 |
| CWA | B2 vs. A2 | 22.333 | 4.749 | [10.070, 34.596] | < 0.001 |
| Study 2 | | | | | |
| Accuracy | G2 vs. G1 | 0.099 | 0.017 | [0.060, 0.138] | < 0.001 |
| Accuracy | G3 vs. G1 | 0.047 | 0.017 | [0.007, 0.086] | 0.015 |
| Accuracy | G3 vs. G2 | −0.053 | 0.017 | [−0.092, −0.013] | 0.005 |
| Precision | G2 vs. G1 | 0.078 | 0.015 | [0.042, 0.114] | < 0.001 |
| Precision | G3 vs. G1 | 0.036 | 0.015 | [−0.000, 0.072] | 0.050 |
| Precision | G3 vs. G2 | −0.042 | 0.015 | [−0.078, −0.006] | 0.017 |
| Recall | G2 vs. G1 | 0.103 | 0.022 | [0.052, 0.154] | < 0.001 |
| Recall | G3 vs. G1 | 0.046 | 0.022 | [−0.005, 0.097] | 0.089 |
| Recall | G3 vs. G2 | −0.057 | 0.022 | [−0.108, −0.006] | 0.025 |
| CWA | G2 vs. G1 | 23.291 | 3.223 | [15.690, 30.892] | < 0.001 |
| CWA | G3 vs. G1 | 12.841 | 3.223 | [5.240, 20.442] | < 0.001 |
| CWA | G3 vs. G2 | −10.450 | 3.223 | [−18.051, −2.849] | 0.004 |

*Notes: Entries report pairwise comparisons from the one-way analyses of variance. Tukey's HSD controls the familywise error rate within each outcome and study. The reported 95% confidence intervals and p values incorporate this correction. The focused B1-A2 Welch contrast reported in Section 4.1.3 is separate from this all-pairs family. Study 2 metrics use cases 16-30; conditions are defined in Section 3.1.*

**Table 2. Collective performance of (synthetic) committees**

| Condition | Individual Average Accuracy | Collective Accuracy $A_c$ | Condorcet Benchmark $P_c$ | Dependence Penalty $d$ | FP | FN |
|---|---|---|---|---|---|---|
| Study 1 | | | | | | |
| A1 | 0.678 | 0.705 | 0.756 | 0.051 | 0.415 | 0.175 |
| A2 | 0.755 | 0.826 | 0.849 | 0.023 | 0.132 | 0.216 |
| B1 | 0.710 | 0.733 | 0.797 | 0.064 | 0.303 | 0.230 |
| B2 | 0.813 | 0.888 | 0.908 | 0.020 | 0.132 | 0.091 |
| Study 2 | | | | | | |
| G1 | 0.778 | 0.821 | 0.874 | 0.052 | 0.161 | 0.191 |
| G2 | 0.877 | 0.932 | 0.958 | 0.026 | 0.066 | 0.069 |
| G3 | 0.824 | 0.861 | 0.919 | 0.058 | 0.148 | 0.133 |

*Notes: In Study 2, all metrics use diagnoses from cases 16–30. FP and FN are conditional on benign and malignant scans, respectively.*

**Figure 1. Performance comparison across groups in Study 1**

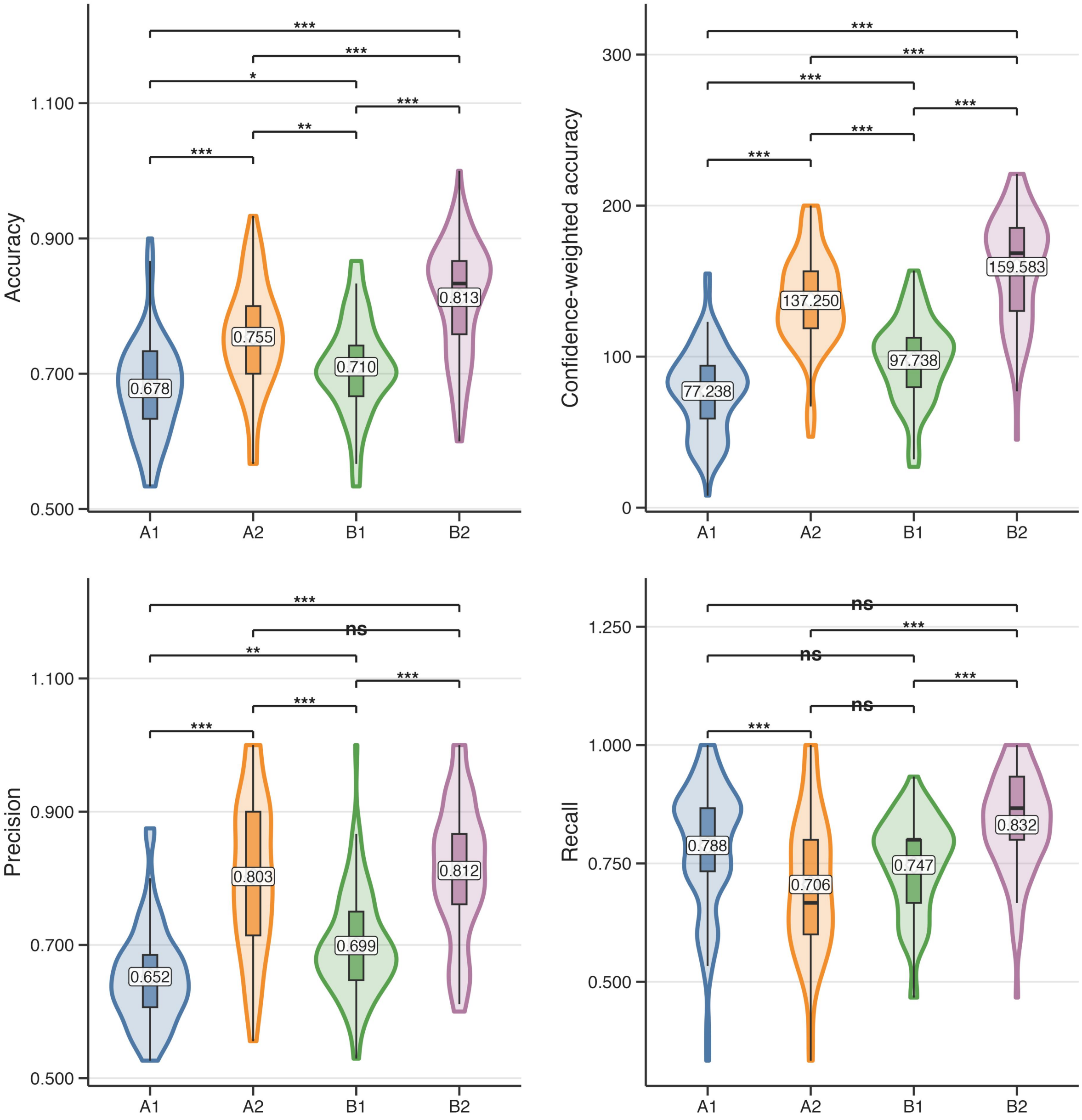

*Note: Values shown inside the violins are condition means. Brackets report the familywise-error-controlled pairwise comparisons in Table 1. Significance codes are † p < 0.100, * p < 0.050, ** p < 0.010, *** p < 0.001, and ns p ≥ 0.100.*

*Group A1: Control condition, in which participants did not receive AI assistance in training or task execution.*

*Group A2: Participants received AI assistance in task execution only.*

*Group B1: Participants received AI assistance in training only.*

*Group B2: Participants received AI assistance in both training and task execution.*

**Figure 2. Distributions of dependence penalty (*d*) in Study 1 and Study 2**

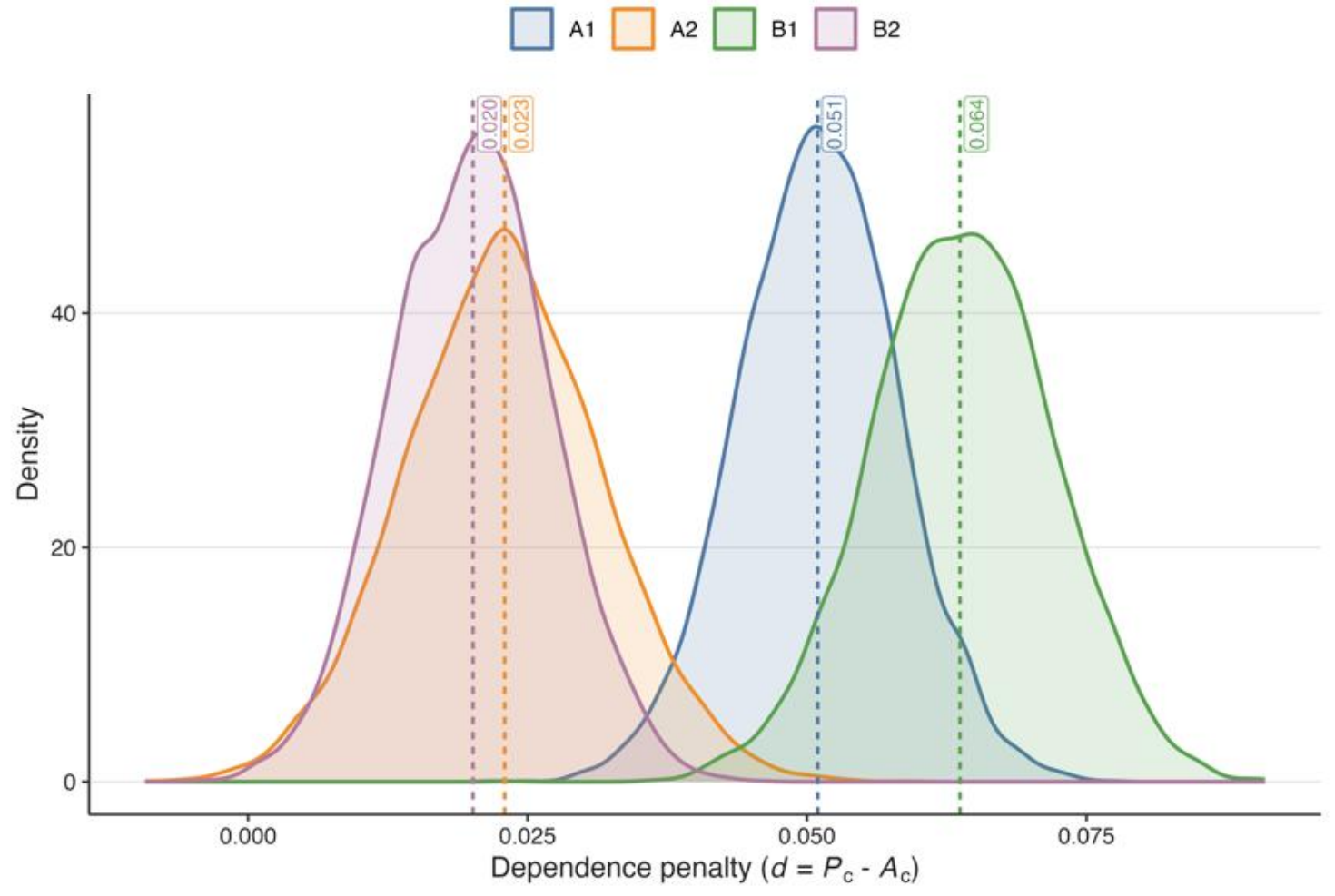

A. Study 1: 28 distinct committees per run, for a total of 5,000 runs

*Note:* A1: Control condition, in which participants did not receive AI assistance in training or task execution.
A2: Participants received AI assistance in task execution only.
B1: Participants received AI assistance in training only.
B2: Participants received AI assistance in both training and task execution.

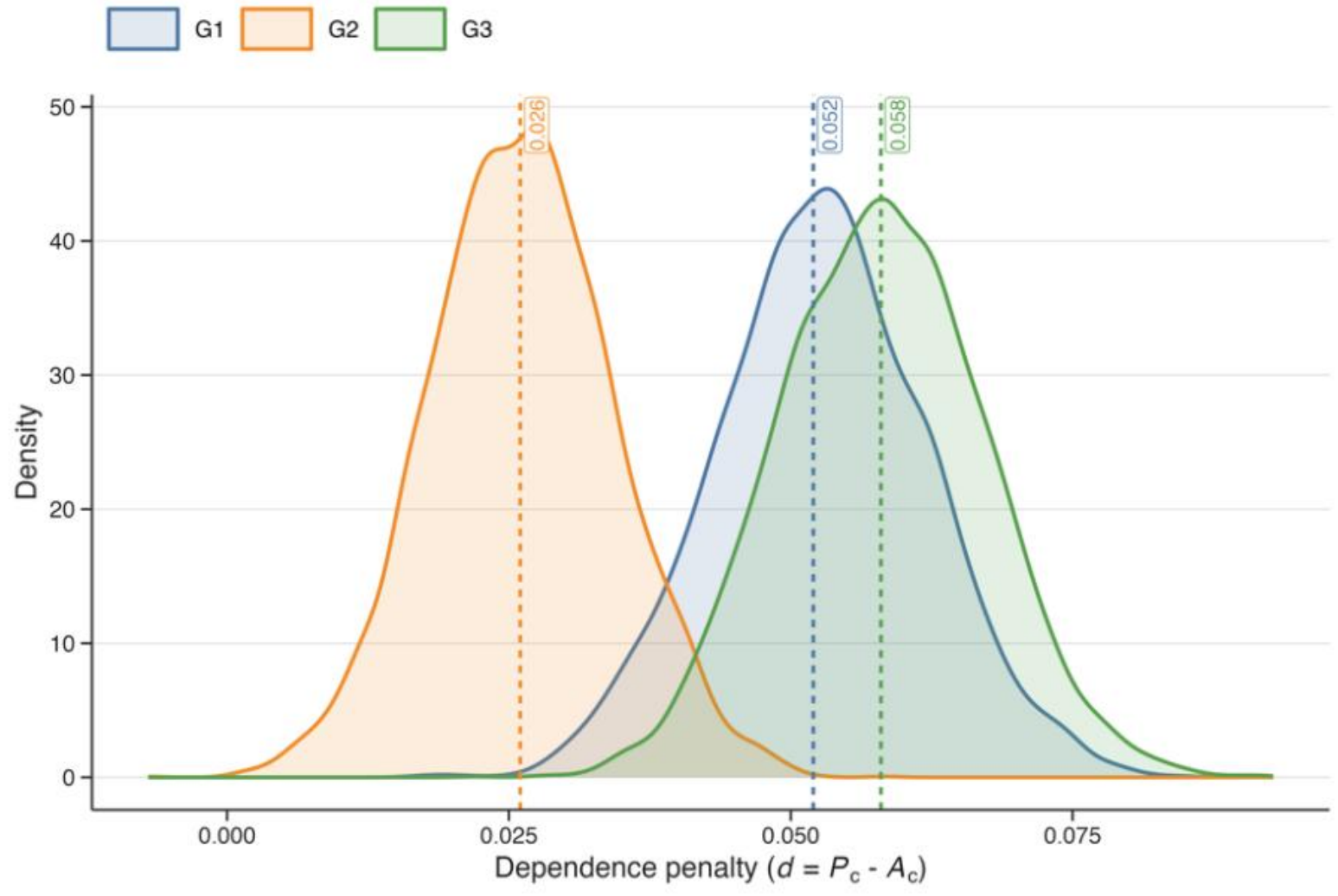

B. Study 2: 26 distinct committees per run, for a total of 5,000 runs

*Note:* Density curves show the 5,000-run Monte Carlo distributions. Dashed vertical lines and labels show the exact all-triad dependence penalties reported in Table 2; these are reference estimates rather than Monte Carlo means. Distributional separation is supported by both Welch two-sample tests (means) and Kolmogorov-Smirnov tests (distributions). All pairwise comparisons are significant at $p < 0.010$.
G1: No AI assistance during either training or task execution.
G2: No AI assistance during training or the second 15 cases, with AI assistance in the first 15 cases.
G3: Reduced AI assistance during training and no AI assistance during task execution.

**Figure 3. Performance comparison across groups in Study 2**

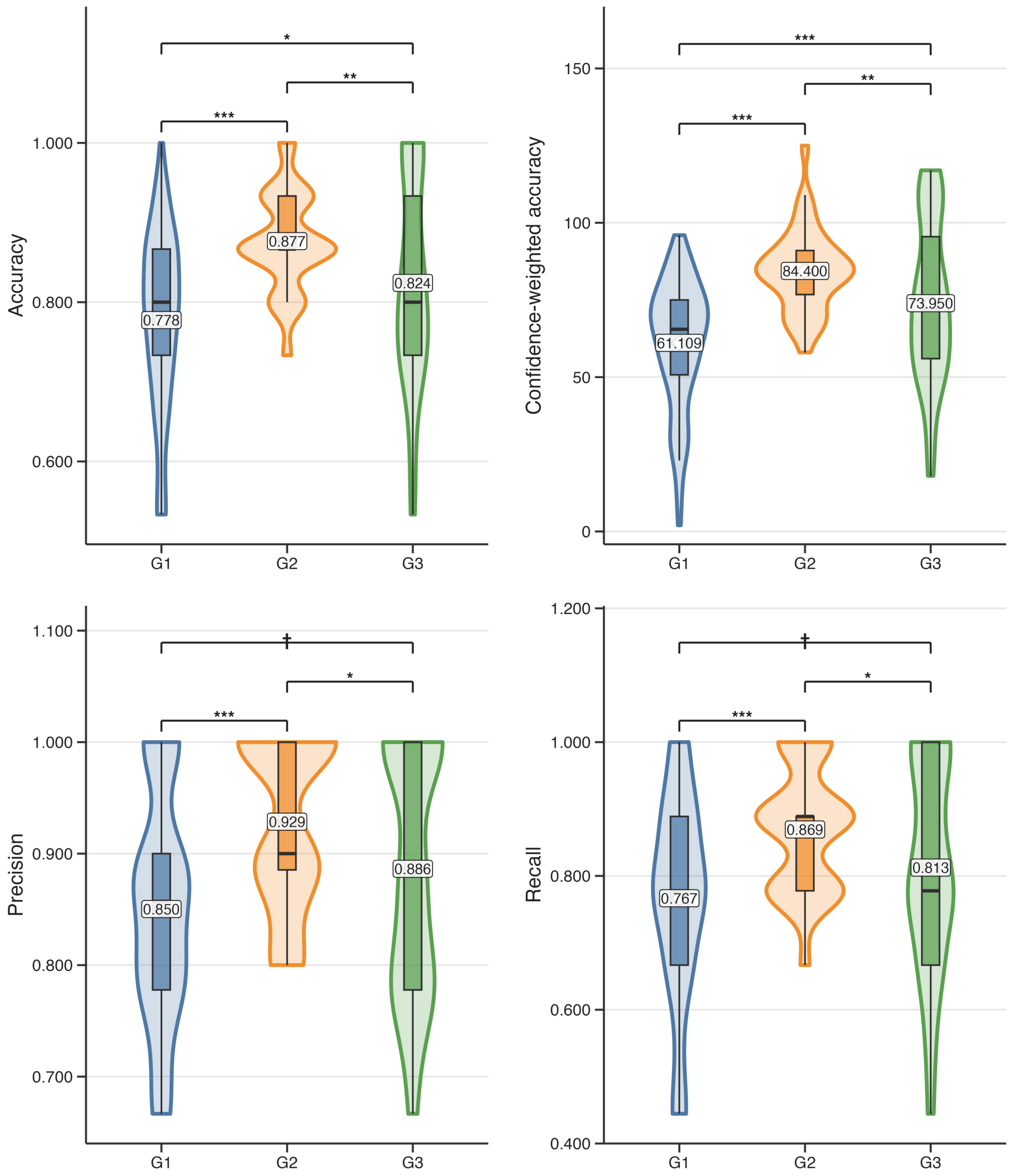

*Note:* Values shown inside the violins are condition means. Brackets report the familywise-error-controlled pairwise comparisons in Table 1. † $p < 0.100$, * $p < 0.050$, ** $p < 0.010$, *** $p < 0.001$, and ns $p \geq 0.100$.

Group 1: No AI assistance during either training or task execution.

Group 2: No AI assistance during training or the second 15 cases, with AI assistance in the first 15 cases.

Group 3: Reduced AI assistance during training and no AI assistance during task execution.

## Supplementary Materials

### Table S1. Age balance test by ANOVA (Study 1)

| Comparison | Mean difference | p value |
|---|---|---|
| Group A1 vs Group A2 | 0.048 | 0.991 |
| Group B2 vs Group A2 | 0.036 | 0.996 |
| Group B1 vs Group A2 | 0.000 | 1.000 |
| Group B2 vs Group A1 | −0.012 | 1.000 |
| Group B1 vs Group A1 | −0.048 | 0.991 |
| Group B1 vs Group B2 | −0.036 | 0.996 |

### Table S2. Gender balance test (Study 1)

| Group | Female n (%) | Male n (%) |
|---|---|---|
| Group A1 | 41 (48.810%) | 43 (51.190%) |
| Group A2 | 41 (48.810%) | 43 (51.190%) |
| Group B1 | 41 (48.810%) | 43 (51.190%) |
| Group B2 | 41 (48.810%) | 43 (51.190%) |

**Table S3. Factorial regression analysis of performance in Study 1**

| | (1) Accuracy | (2) Accuracy | (3) CWA | (4) CWA | (5) Precision | (6) Precision | (7) Recall | (8) Recall |
|---|---|---|---|---|---|---|---|---|
| AI assistance in task execution | 2.298*** | 2.312*** | 60.012*** | 60.090*** | 0.150*** | 0.151*** | −0.083*** | −0.082*** |
| | (0.354) | (0.355) | (4.749) | (4.756) | (0.014) | (0.014) | (0.019) | (0.019) |
| AI assistance in training | 0.964** | 0.974** | 20.500*** | 20.633*** | 0.047** | 0.047*** | −0.041* | −0.041* |
| | (0.354) | (0.354) | (4.749) | (4.750) | (0.014) | (0.014) | (0.019) | (0.019) |
| Task-execution assistance × training assistance | 0.786 | 0.765 | 1.833 | 1.655 | −0.038† | −0.039† | 0.167*** | 0.167*** |
| | (0.501) | (0.502) | (6.716) | (6.722) | (0.020) | (0.020) | (0.027) | (0.027) |
| Age | | 0.196 | | 2.802† | | 0.012* | | −0.003 |
| | | (0.123) | | (1.646) | | (0.005) | | (0.007) |
| Female | | −0.009 | | −1.313 | | −0.006 | | 0.005 |
| | | (0.252) | | (3.375) | | (0.010) | | (0.014) |
| Prior AI diagnostic experience | | −0.069 | | 0.780 | | 0.003 | | −0.008 |
| | | (0.260) | | (3.487) | | (0.010) | | (0.014) |
| Constant | 20.345*** | 15.913*** | 77.238*** | 13.678 | 0.652*** | 0.374*** | 0.788*** | 0.846*** |
| | (0.250) | (2.807) | (3.358) | (37.625) | (0.010) | (0.113) | (0.013) | (0.151) |
| N | 336 | 336 | 336 | 336 | 336 | 336 | 336 | 336 |
| $R^2$ | 0.307 | 0.313 | 0.527 | 0.531 | 0.352 | 0.365 | 0.128 | 0.129 |

*Notes: Standard errors are in parentheses. Accuracy is the number of correct diagnoses out of 30; CWA is confidence-weighted accuracy. Columns (2), (4), (6), and (8) include age, female, and prior AI diagnostic experience. † p < 0.100; * p < 0.050; ** p < 0.010; *** p < 0.001.*

**Table S4. Divergence from AI prediction in Group A2 and Group B2 (Study 1)**

| AI classification | A2 mean divergence | B2 mean divergence | B2−A2 | s.e. | 95% CI | t | d.f. | p |
|---|---|---|---|---|---|---|---|---|
| Correct (0.500 cutoff) | 0.228 | 0.155 | −0.072 | 0.012 | [−0.096, −0.049] | −5.985 | 4,198 | < 0.001 |
| Incorrect (0.500 cutoff) | 0.667 | 0.655 | −0.012 | 0.033 | [−0.076, 0.052] | −0.364 | 838 | 0.716 |
| Correct (0.950 cutoff) | 0.207 | 0.186 | −0.021 | 0.015 | [−0.050, 0.008] | −1.412 | 2,854 | 0.158 |
| Incorrect (0.950 cutoff) | 0.705 | 0.812 | 0.107 | 0.018 | [0.071, 0.143] | 5.895 | 2,182 | < 0.001 |

*Note: AI probabilities at or above the specified malignancy-probability cutoff were classified as malignant and lower probabilities as benign. At the 0.500 and 0.950 cutoffs, 5 and 13 of the 30 classifications were incorrect, respectively. Divergence was analyzed at the participant–case level; d.f. = degrees of freedom. Statistics were calculated using unrounded values; displayed values are rounded to three decimal places.*

*Group A2: Participants received AI assistance in task execution phase only.*

*Group B2: Participants received AI assistance in both training and task execution phases.*

**Table S5. Age balance test by ANOVA (Study 2)**

| Comparison | Mean difference | *p* value |
|---|---|---|
| Group 2 vs Group 1 | −0.063 | 0.928 |
| Group 3 vs Group 1 | −0.150 | 0.651 |
| Group 3 vs Group 2 | −0.088 | 0.864 |

**Table S6. Gender balance test (Study 2)**

| Group | Female *n* (%) | Male *n* (%) |
|---|---|---|
| Group 1 | 39 (48.750%) | 41 (51.250%) |
| Group 2 | 39 (48.750%) | 41 (51.250%) |
| Group 3 | 42 (52.500%) | 38 (47.500%) |

**Figure S1. Research procedure for Study 1**

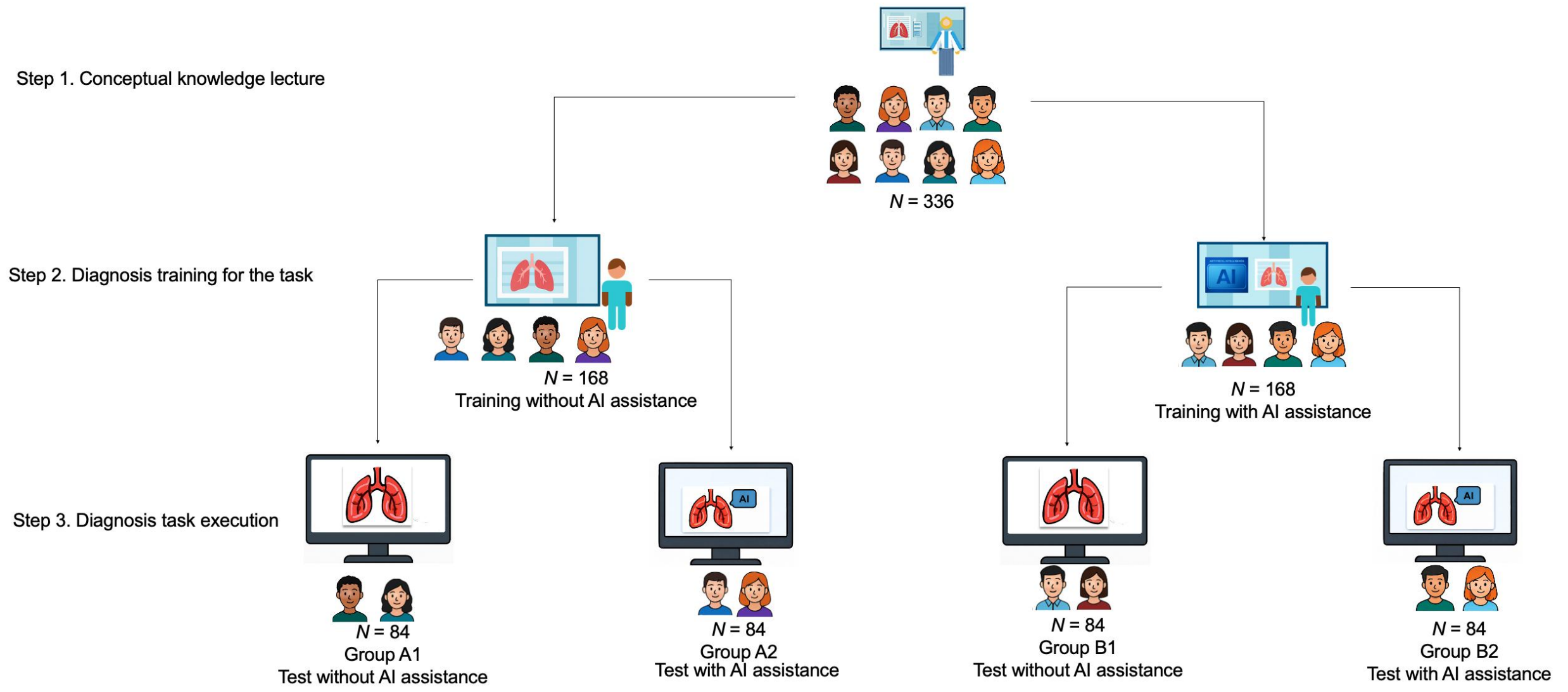

**Figure S2. Research procedure for Study 2**

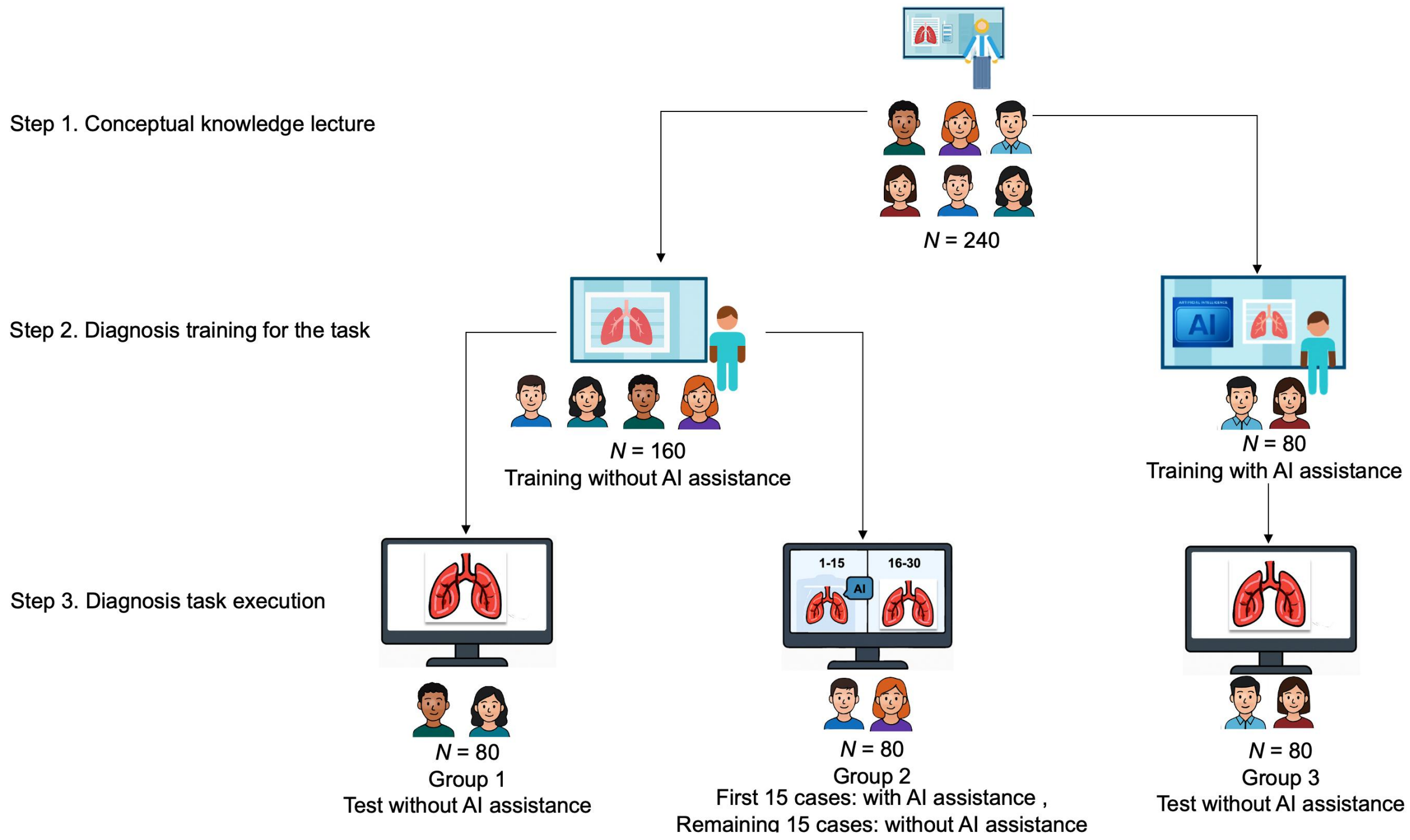

**Figure S3. Two types of AI assistance deployed in our experiments**

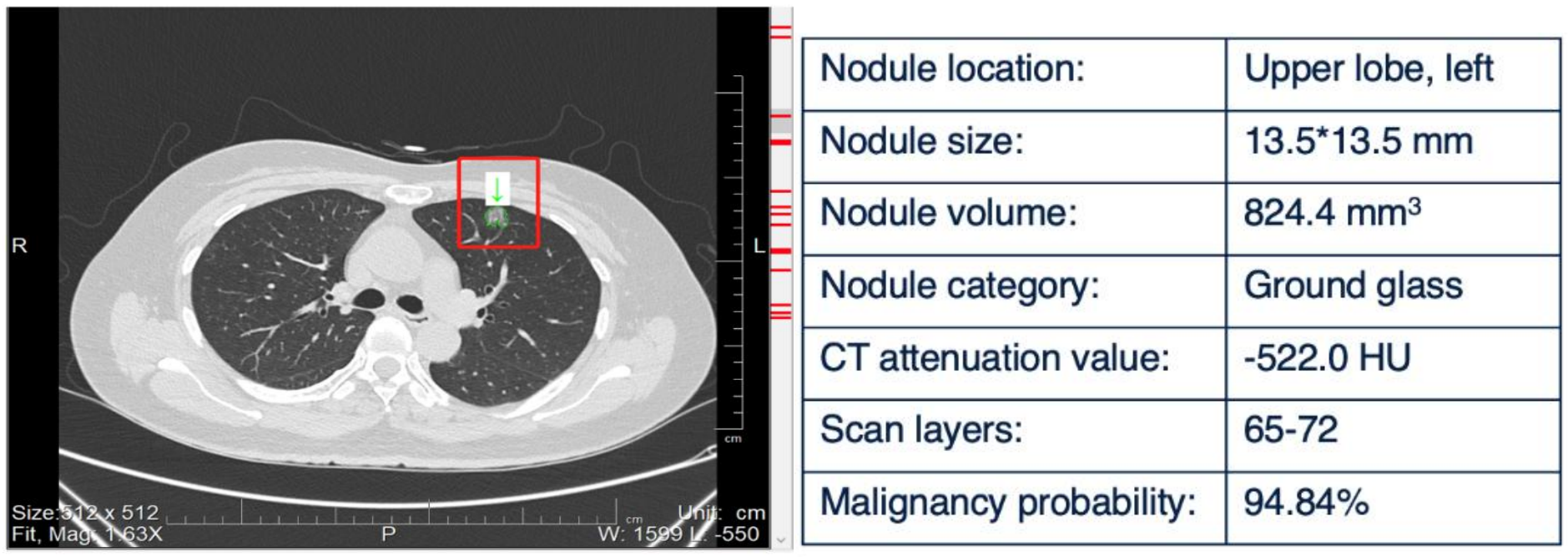

**A. Full AI input (Study 1: Group B1 and B2 in training; Study 2: Group 2 in task execution)**

AI assistance included (1) a green highlight around the identified nodule, (2) a predicted malignancy probability (0–100%), and (3) seven quantified diagnostic features.

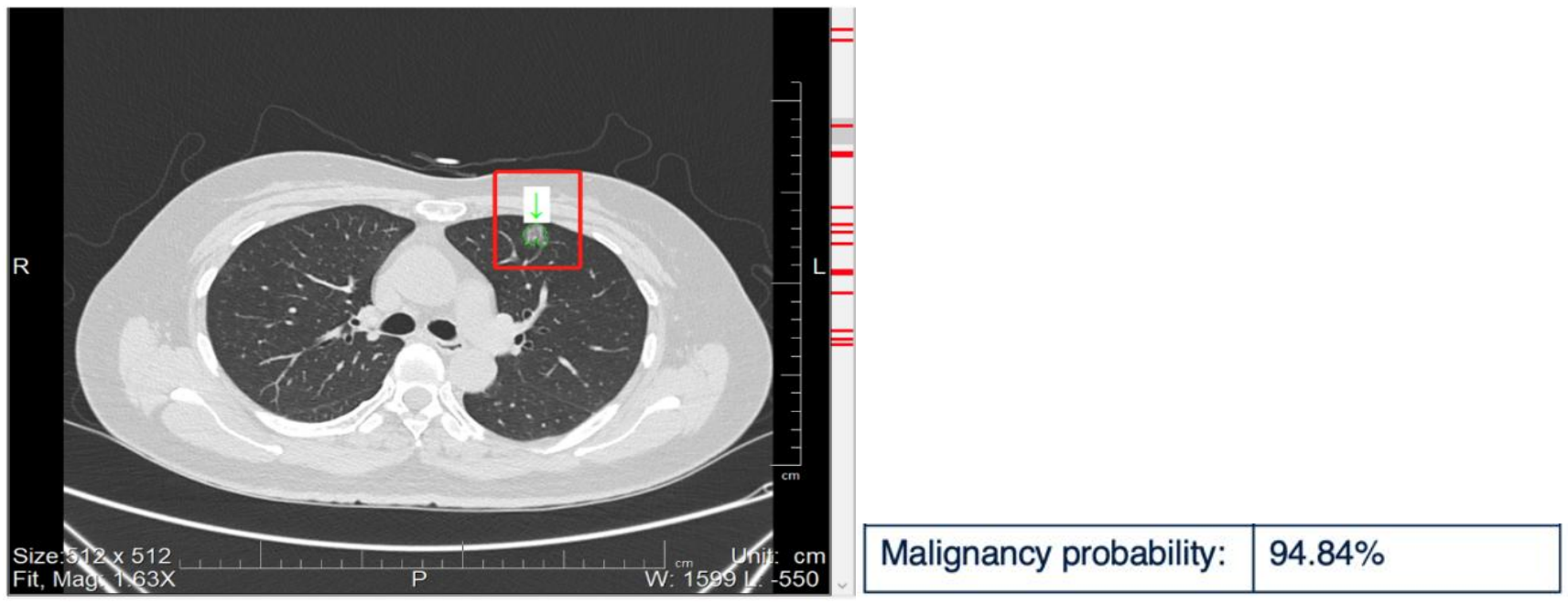

**B. Reduced AI input (Study 2: Group 3 in training)**

The AI system provided only the identified nodule and a predicted malignancy probability (0–100%).

**Figure S4. Receiver-Operating Characteristic Curve (ROC) of the predictive AI system**

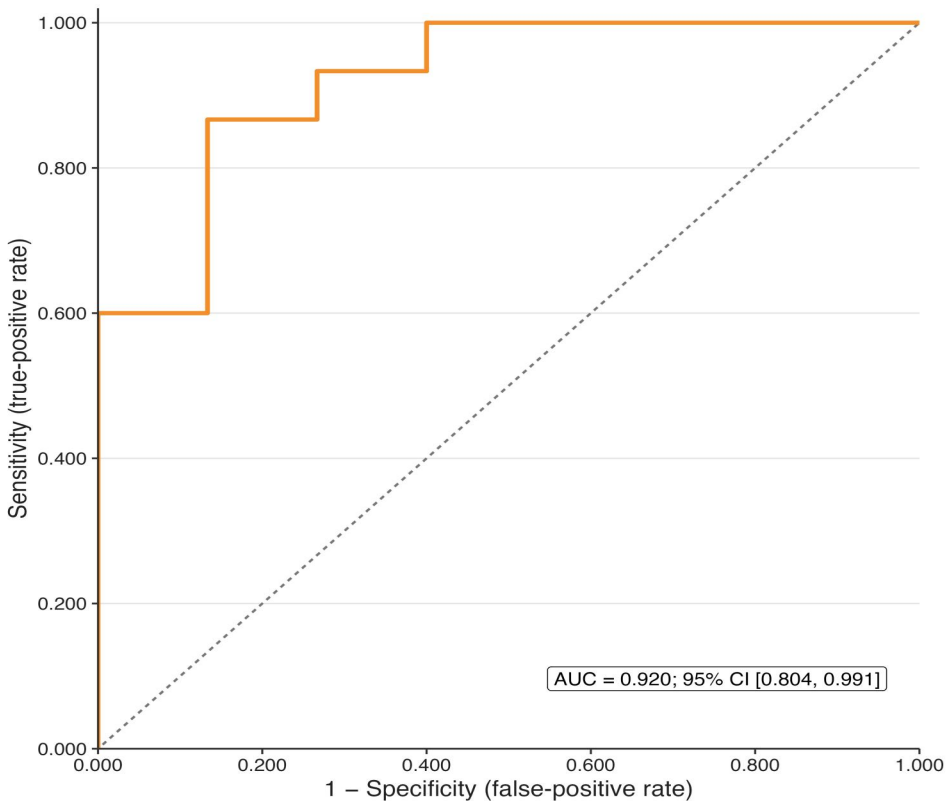

**Figure S5. Calibration scores across conditions (Study 1 and Study 2)**

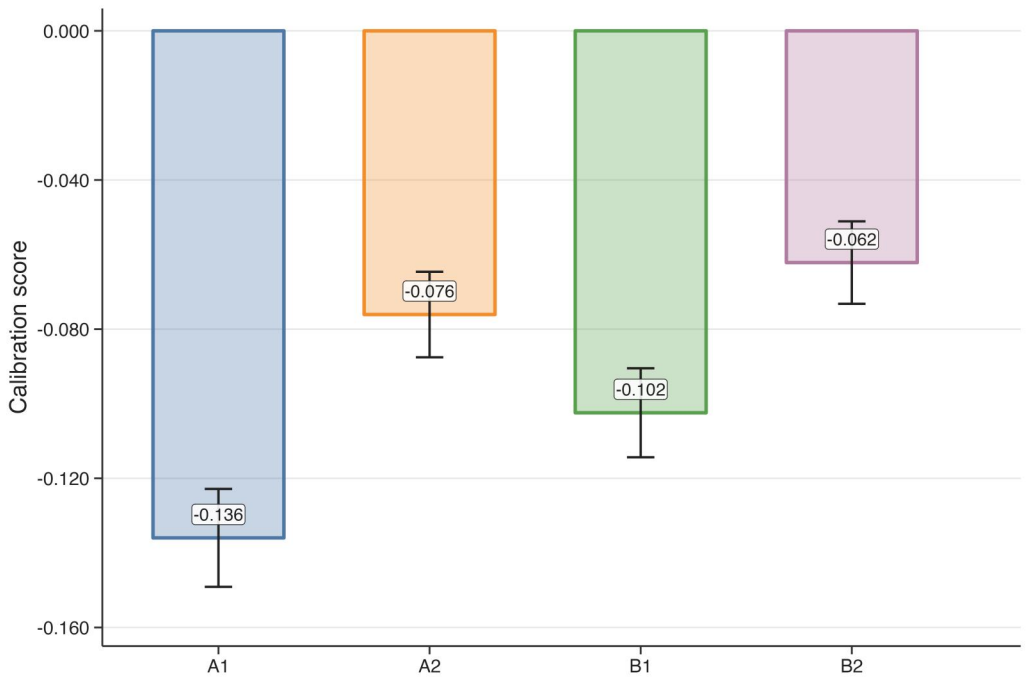

Study 1

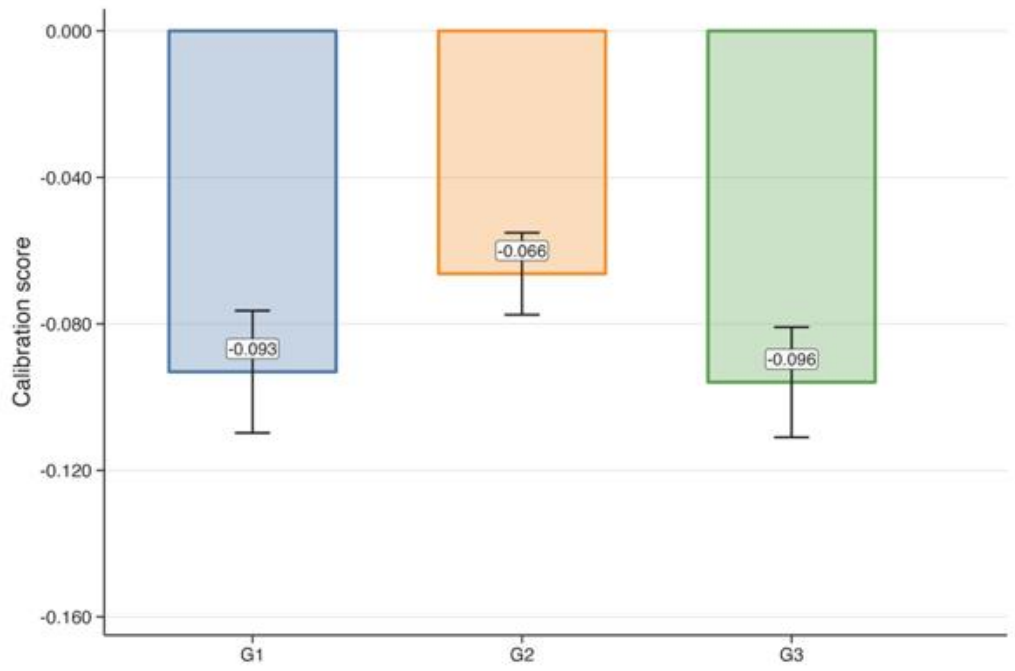

Study 2

*Note: Bars show mean participant-level calibration scores; error bars are 95% confidence intervals. Values closer to 0.000 indicate better calibration.*

**Figure S6. Comparison of common error index across conditions (Study 1 and Study 2)**

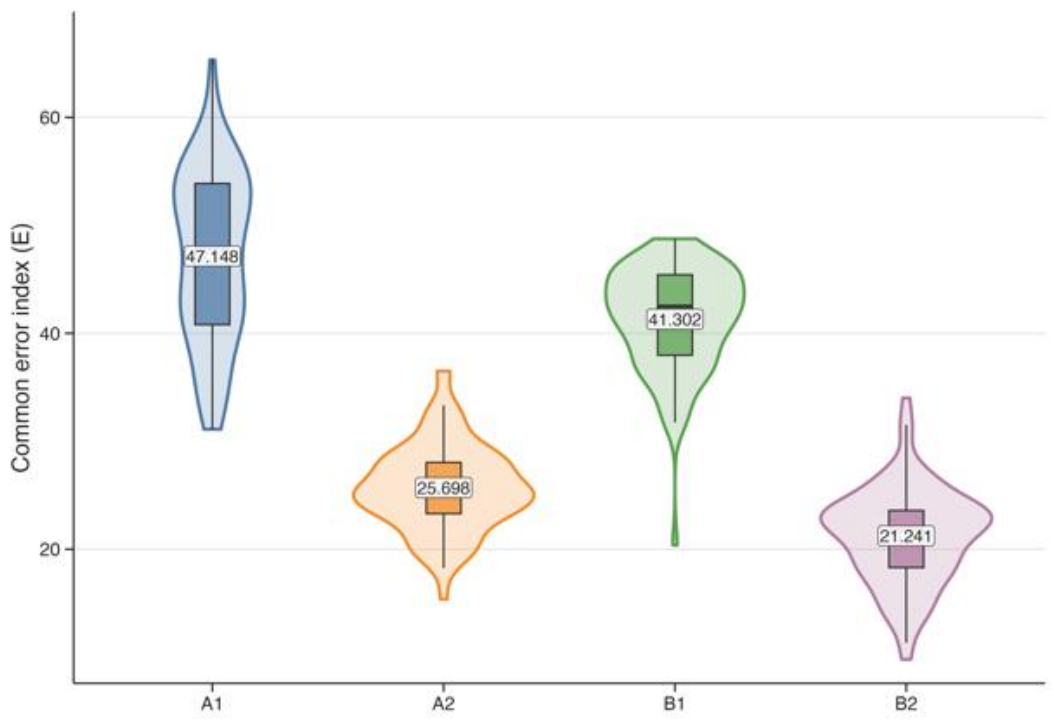

Study 1

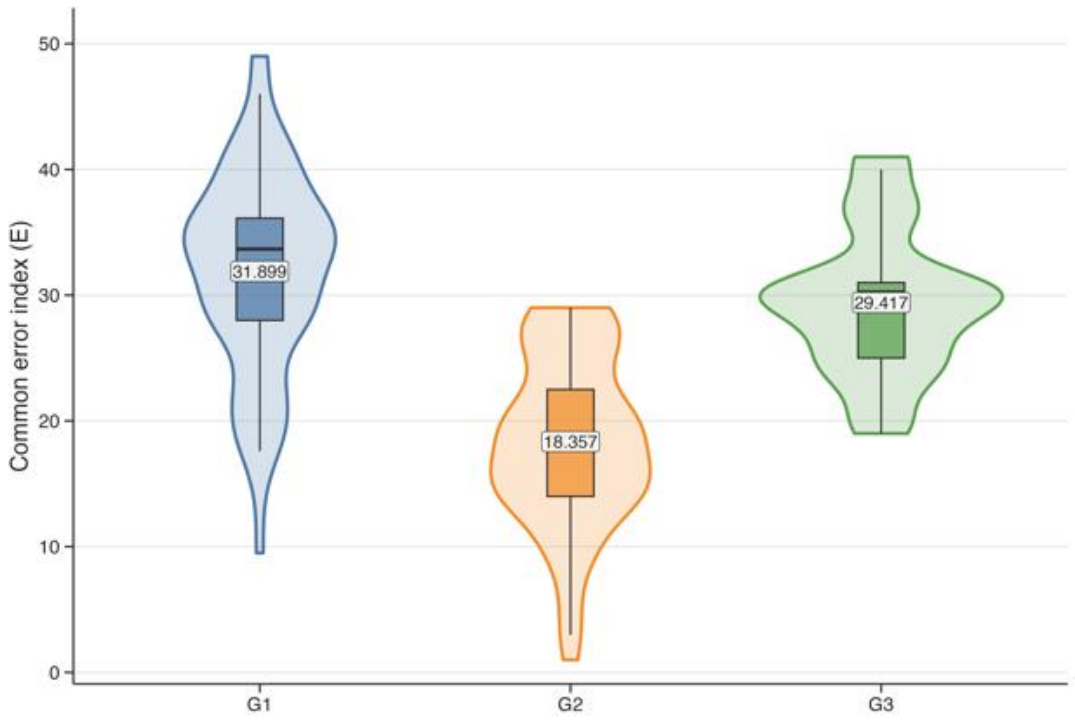

Study 2

*Note: The common-error index was calculated using all 30 scans in Study 1 and cases 16–30 in Study 2. Pairwise Welch tests in Study 1 all had $p < 0.001$. In Study 2, G1 versus G2 and G2 versus G3 had $p < 0.001$; G1 versus G3 had $p = 0.034$.*